\documentclass[12pt,preprint]{aastex}

\newcommand{\masyr}{${\rm ~mas~yr}^{-1}$}

\shorttitle{Lick NPM2 Catalog}
\shortauthors{Hanson et al.}

\begin{document}

\title{Lick Northern Proper Motion Program. III.\\
Lick NPM2 Catalog}

\author{Robert B. Hanson, Arnold R. Klemola, and Burton F. Jones} 
\affil{University of California Observatories/Lick Observatory,\\
University of California, Santa Cruz, CA 95064}
\email{hanson,klemola,jones@ucolick.org}

\and
\author{David G. Monet} 
\affil{US Naval Observatory, Flagstaff Station, P.O. Box 1149,
Flagstaff, AZ 86002}
\email{dgm@nofs.navy.mil}

\begin{abstract} 
The Lick Northern Proper Motion (NPM) program, a two-epoch (1947--1988)
photographic survey of the northern two-thirds of the sky ($\delta \gtrsim 
-23\degr$), has measured absolute proper motions, on an inertial system defined
by distant galaxies, for 378,360 stars from $8 \lesssim B \lesssim 18$.  
The 1993 NPM1 Catalog contains 148,940 stars in 899 fields outside the 
Milky Way's zone of avoidance. The 2003 NPM2 Catalog
contains 232,062 stars in the remaining 347 NPM fields near the plane of the 
Milky Way.  This paper describes the NPM2 star selection, plate measurements,
astrometric and photometric data reductions, and catalog compilation. 
The NPM2 Catalog contains 122,806 faint ($B \geq 14$) anonymous stars for
astrometry and galactic studies, 91,648 bright ($B < 14$) positional reference 
stars, and 34,868  ``special stars'' chosen for astrophysical interest.  
The NPM2 proper motions are on the ICRS system, via Tycho-2 stars, to an 
accuracy of 0.6\masyr\ in each field.  RMS proper motion precision is 6\masyr.
Positional errors average 80~mas at the mean plate epoch 1968, and 200~mas at 
the NPM2 catalog epoch 2000.  NPM2 photographic photometry errors average 
0.18~mag in $B$, and 0.20~mag in \bv.  The NPM2 Catalog and the updated 
(to J2000) NPM1 Catalog are available at the CDS Strasbourg data center and on
the NPM WWW site\footnote{\tt http://www.ucolick.org/${\thicksim}$npm}. 
The NPM2 Catalog completes the Lick Northern Proper Motion program after a 
half-century of work by three generations of Lick Observatory astronomers.  
The NPM Catalogs will serve as a database for research in galactic structure, 
stellar kinematics, and astrometry.

\end{abstract}


\setcounter{table}{0}

\section{Introduction} \label{INT}

The study of the motions of various classes of stars is fundamental to 
an understanding of our Galaxy.  Proper motions provide the observational 
basis for the study of solar motion, galactic rotation, and stellar velocity 
dispersions, and for the statistical determination of stellar distances 
and luminosities.  The Lick Northern Proper Motion (NPM) program
\citep[][Paper~I]{kjh87}, a two-epoch photographic survey of the northern 
two-thirds of the sky, now complete after a half-century of work, meets many 
of these basic observational needs by measuring absolute proper motions
on an inertial system defined by distant galaxies, accurate positions,
and $BV$ photographic photometry for 378,360 stars over a blue apparent 
magnitude range from 8 to 18.  

The NPM program includes many types of stars chosen for astrophysical interest,
anonymous stars for astrometry and galactic studies, and a selection of stars 
from positional catalogs and from other proper motion surveys.
The NPM survey consists of $1,246\ \ 6\degr \times 6\degr$ fields with 
$\delta \gtrsim -23\degr$ photographed with the Lick 51~cm (20~in) Carnegie 
double astrograph (scale $\sim$55$\,\farcs 1$~mm$^{-1}$) at two epochs 
(1947--1954 and 1970--1988).  Paper~I describes the Lick astrograph and plate 
material, and gives a comprehensive overview of the history, goals, and methods
of the NPM program.

\begin{figure}
\includegraphics[scale=0.565]{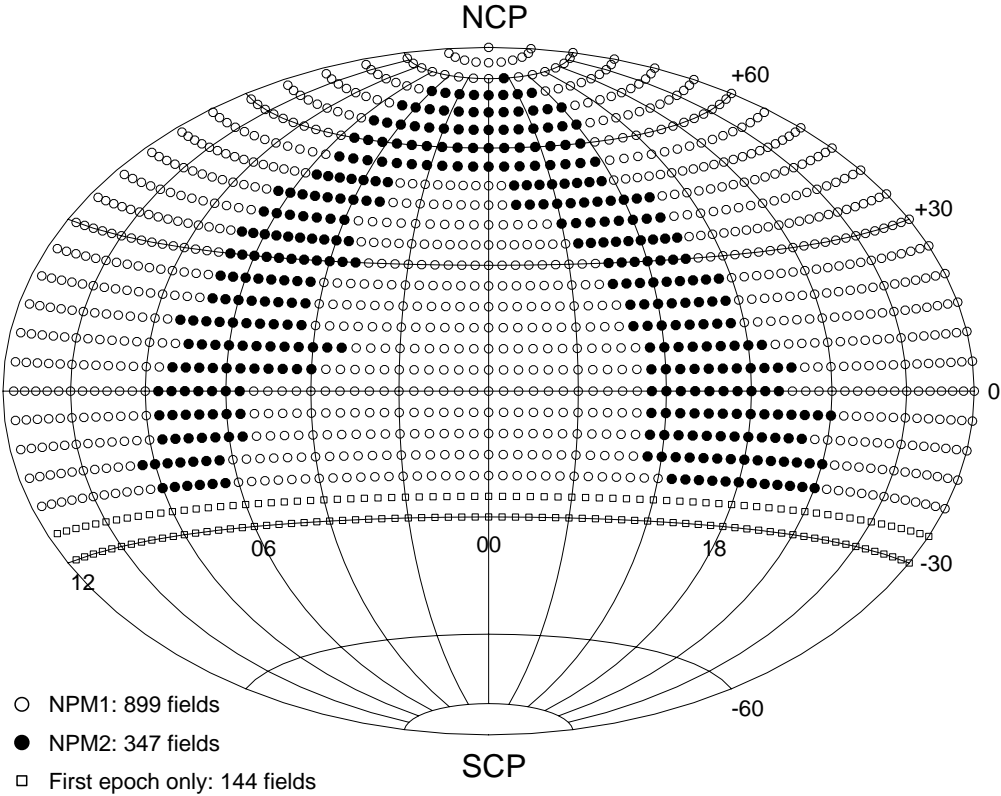}
\caption[fig1.eps]{
Hammer-Aitoff plot showing the distribution of NPM field centers on the sky.  
Coordinates are B1950 right ascension and declination.  
The Lick astrograph plates cover $6\fdg3 \times 6\fdg3$.  
Fields are centered in $5\degr$ declination bands.  
Right ascension spacing is $20^{\rm m}$ ($5\degr$ at the Equator) for 
$\vert\delta\vert \leq 35\degr$, and is progressively adjusted for 
$\delta \geq 40\degr$ so that neighboring fields overlap by at least $1\degr$.
The 1993 NPM1 Catalog comprises 899 fields (open circles) with
$-20\degr \leq \delta \leq +90\degr$ away from the Milky Way.  
The NPM2 Catalog comprises 347 fields (filled circles) with 
$-20\degr \leq \delta \leq +80\degr$ at low galactic latitudes
($\vert b \vert \lesssim 15\degr$).  Open squares show the 144
``Southern Extension'' fields at $-25\degr$ and $-30\degr$, which were
photographed at first epoch only, and are not included in the NPM program.
\label{skyNPM} 
}
\end{figure}

The main results of the Lick NPM program (positions, proper motions, 
magnitudes, and colors) are published in two catalogs, the NPM1 Catalog 
\citep{khj93a} with 148,940 stars in 899 ``NPM1'' fields outside the Milky 
Way's zone of avoidance, and the NPM2 Catalog \citep{hkjm03}, with 232,062 
stars in the remaining 347 ``NPM2'' fields near the plane of the  Milky Way.  
Figure~\ref{skyNPM} shows the sky distribution of the NPM1 and NPM2 fields.
As a service to users, the 1993 NPM1 Catalog has been updated \citep{khj00} 
from the obsolete B1950 coordinates to the J2000 epoch and equinox
used for NPM2.  The two NPM Catalogs and associated documentation files are 
available from the CDS Strasbourg data center and on the Lick NPM program's 
WWW site ~ {\tt http://www.ucolick.org/${\thicksim}$npm}.

Several products supplementing the NPM Catalogs have also been compiled.  
The Lick Input Catalog of Special Stars (ICSS; see Paper~I) represents
Klemola's continuing search of the astronomical literature since 1972
to include many classes of stars of astronomical or astrophysical interest 
in the NPM program.  Klemola's ICSS now contains some 300,000 entries for
220,000 stars from over 1,100 literature references, and forms the basis for 
two published catalogs.  The NPM1 Cross-Identifications \citep{khj94} contains
41,858 entries with source identifications, stellar type classifications, 
publication references, footnotes, and other information to assist users of 
the NPM1 Catalog.  The NPM2 Cross-Identifications \citep{khjm04}, with 46,887 
entries, is the corresponding supplement to the NPM2 Catalog.  
Another supplement to NPM1 is the NPM1 Reference Galaxies list \citep{khj93b},
giving $0\,\farcs2$ precision positions and $B$ magnitudes for 50,517 faint 
galaxies (mostly $16 \lesssim B \lesssim 18$) used as the proper motion 
reference frame for NPM1.  

The ultimate goal of the NPM program, in coordination with the Yale Southern
Proper Motion (SPM) program \citep{gir98,plx98a,gir04} is to provide accurate 
absolute proper motions for faint stars ($B \gtrsim 12$) over the whole sky, 
on a reliable, uniform system, linked \citep{kov97,plx98b} to the 
ICRS/Hipparcos \citep{esa97} proper motion frame realized by bright stars 
($B \lesssim 12$).  
The scientific significance of the Lick NPM program is that
its large size, extensive magnitude range $(8 \lesssim B \lesssim 18)$, 
high accuracy, and absolute proper motions will ensure its lasting value
as a unique database for research in galactic structure and kinematics 
\citep{han87,bex00}, stellar luminosities \citep{lay96} and 
astrometry \citep{plx98b}.

\citet{khj95} and \citet{han97} have described the NPM1 Catalog and its
applications.  This paper will describe the NPM2 Catalog, including
details of the NPM2 star selection, astrometric and photometric data
reductions, and catalog compilation.  We will emphasize the differences
between NPM2 and NPM1, and compare the characteristics of both catalogs. The
reader should be aware that the NPM2 plate material, the general structure of
the astrometric and photometric data reduction ``pipeline'', and much of the
NPM2 catalog compilation procedures and error tests are essentially similar
to NPM1.  Details not mentioned in this paper are as described in Paper~I
(plate material, reduction procedures), or in \citet{khj95} and \citet{han97}
(catalog compilation and contents). The most important differences between
NPM2 and NPM1 are:

\begin{enumerate}

\item
For NPM2, the Lick astrograph plates were scanned by the Precision Measuring
Machine (PMM) at the US Naval Observatory, Flagstaff (Section~\ref{PMM}).
Stars selected for NPM2 were identified and extracted from the PMM scans
(Section~\ref{STS}).
For NPM1, the plates were measured on the Automatic Measuring Engine (AME) 
at Santa Cruz, using an input list of stars previously selected at the Lick 
Survey Machine.

\item
With the PMM scans, NPM2 could use many more reference stars than NPM1 
in the astrometric and photometric data reductions (Sections~\ref{POS},
\ref{PMS}, and \ref{PHOT}).
This allowed more rigorous plate constant models and more robust reductions.

\item
NPM2 fields lie in the zone of avoidance, and have far too few galaxies to 
determine the zero-point corrections from relative to absolute proper motion 
as per NPM1.  For NPM2, we used Tycho-2 Catalogue \citep{hog00a,hog00b} 
stars to correct each field's relative proper motions to the ICRS system
(Section~\ref{PMS}).

\end{enumerate}

Section~\ref{SSPM} describes how the different classes of stars were selected
for NPM2, briefly discusses the PMM scans of the NPM plates, and explains how
the $X,Y$ and magnitude measurements for NPM2 stars were extracted from the 
PMM scans.  Section~\ref{RED} outlines the NPM2 data reductions,
describes the selection of astrometric reference stars, and discusses the
astrometric modeling of the Lick astrograph plates.  Sections~\ref{POS}, 
\ref{PMS}, and \ref{PHOT} explain the NPM2 astrometric (positions and proper 
motions) and photometric ($B$~magnitude, \bv~color) reductions and internal 
error tests.
Section~\ref{CAT} tells how the NPM2 Catalog was compiled.  Multiple measures 
from the 347 overlapping fields were averaged, checked for discordances, and 
used to estimate the accuracy of the NPM2 positions, proper motions, and 
photometry.  We outline the organization of the NPM2 Catalog, and discuss the 
NPM2 Cross-Identifications.
Finally, Section~\ref{SUM} summarizes the NPM program and outlines present
or potential applications of the NPM data to galactic research.

\section{Star Selection and Plate Measurement} \label{SSPM}

For NPM1, the selection of stars in each field and the measurement of the
plates were two separate processes, described in Section~IV of Paper~I.  
NPM2 is more complicated.  For the most part, broad categories 
of stars desired for NPM2 were defined in advance, but the specific stars were 
chosen {\it a posteriori} when the data were extracted from the PMM scans. 

\subsection{NPM2 Stellar Content} \label{SCON}

The stellar content of the NPM1 Catalog, including Klemola's ICSS, 
was detailed in Paper~I.  The content of the NPM2 Catalog is broadly similar 
to NPM1, with the major additions discussed  below, and other differences 
described by \citet{han97}, most notably Klemola's continued updating of the 
ICSS. Six broad categories of stars selected for the NPM program are outlined 
in Table~\ref{tbl-1}.
NPM2 selected about four times as many stars per field as NPM1. The main
reasons are:

\begin{enumerate}

\item
To improve our proper motion reductions (Section~\ref{PMS}), and to provide 
a denser network (10~stars~deg$^{-1}$) of faint ($B \gtrsim 16$) stars with 
accurate positions, NPM2 selected over 400 faint anonymous stars per field, 
compared to 72 or 144 for NPM1.

\item
NPM2 includes large numbers of stars from Hipparcos, Tycho, and other catalogs
not available to NPM1.  Most notably, NPM2 includes $\sim$400 astrometric 
reference stars per field from the 2.5-million-star Tycho-2 Catalogue 
\citep{hog00a,hog00b}, greatly improving the accuracy of our positional
reductions (Section~\ref{POS}).

\item
Many classes of stars are concentrated to the galactic plane or toward the 
galactic center, so the average sky density for NPM2 is much higher than for 
NPM1.  About 55\% of Klemola's ICSS, some 123,000 stars, fall into the 347 
NPM2 fields.\footnote{As discussed in Section \ref{SIC}, only about 30\% of 
these ICSS stars ``survive'' in the NPM2 Catalog.  Star counts cited here
are ICSS counts.  Section \ref{CROSS} gives NPM1~$+$~NPM2 Catalog counts 
for several major stellar classes.}
Abundant ICSS classes that show strong concentration to the Milky Way include 
7000 OB stars, 5000 red giants, 3000 carbon stars, and 2000 RR Lyrae type
variable stars.

\item
Klemola's literature surveys continually added stars to the ICSS, which is now
four times as large as the version used for NPM1.  Table~II of Paper~I lists 
the major classes of stars in Klemola's ICSS.  The ICSS also includes large 
numbers of stars from two-color and objective-prism surveys, and many stars 
from recent surveys of the galactic anticenter and elsewhere.  Several major 
catalogs or compilations are now included whole in the ICSS.  These include 
the GCVS4 \citep{gcvs} variable star catalog with its supplements, the new 
Yale Parallax Catalogue \citep{ypc}, the Third Catalogue of Nearby Stars 
\citep{cns}, and the NLTT proper motion catalog \citep{nltt}. These additions 
will greatly enhance the body of basic data for galactic studies.

\end{enumerate}

In total, the NPM2 Catalog (Section~\ref{CAT}) contains 122,806 faint anonymous
stars selected for the NPM2 astrometric reductions and for statistical studies 
of stellar motions; 91,648 positional reference stars, mostly from the Tycho-2
Catalogue; and 34,868  ``special stars'' from Klemola's ICSS.
These categories overlap slightly; the final NPM2 Catalog (Section~\ref{CAT})
totals 232,062 stars from all 347 NPM2 fields.  

\subsection{PMM Scans} \label{PMM}

From September 1996 to January 1999, under the supervision of Monet, a total of
$3,882\ \ 17 \times 17$~in plates, essentially the entire NPM plate collection,
was scanned with the Precision Measuring Machine \citep[PMM;][]{mon03} at the 
US Naval Observatory, Flagstaff.  This includes the 899 fields previously
measured at Lick for NPM1, as well as the 347 NPM2 fields.  There are three
plates per NPM field: blue at first epoch (``B1''), blue (``B2'') and 
yellow (``Y2'') at second epoch. (For completeness, PMM also scanned 144 B1
plates in the two Southern Extension zones (centers at $\delta = -25\degr$ 
and $-30\degr$), which were photographed at Lick in the 1950s but not repeated 
for the NPM program.)
The operation of the PMM and the reduction of the pixel data (transmission
values) to $X,Y$ coordinates and instrumental magnitudes ($m_b$ or $m_y$)
were as described by \citet{gir04} for the SPM program.
Plates were scanned in the ``direct'' orientation only.

Positional and photometric data for all images detected (up to $3.4 \times 10^6$
per plate) were saved on 207 CD-ROMs, comprising $\sim$130~Gb of data for
further analysis.  Data for the 347 NPM2 fields were extracted for astrometric
and photometric reductions.  Many other uses of the NPM plate scan database
are possible, such as expanding the NPM1 Catalog, or making all-sky catalogs 
beyond the limited scope of the NPM program \citep{mon03}.

The completion of the NPM plate scans at Flagstaff, with a new generation 
of measuring machine, truly represents a remarkable increase in speed; 
by comparison, it took 17 years to survey and measure the 899 NPM1 fields 
on the Lick AME at Santa Cruz. 

The NPM2 plate reductions (Secs.~\ref{POS}, \ref{PMS}, and \ref{PHOT}) 
verified that the positional (1--2~$\mu$m $\simeq 50$--100~mas) and 
photometric ($\sim$0.2~mag) precision of PMM on the NPM plates are
quite comparable to the Lick AME 
(Paper~I) when oversize, weak, or blended images are excluded from the data 
reduction pipeline, which we took considerable effort to do for NPM2.  
This allows the NPM2 Catalog, from PMM scans, to be as precise as 
what we achieved for NPM1 from AME measurements.

\subsection{NPM2 Star Image Data Extraction} \label{STS}

The main purpose of the NPM program (Paper~I) is to establish a reference
frame and determine absolute proper motions for stars of astrophysical or
kinematical interest, not to measure every star on the plates.  
Here we describe how we selected from the PMM scans (averaging $\sim 10^6$ 
image detections per NPM2 plate) those stars (and image combinations) 
we wanted for NPM2.
This task is complicated by the fact that the Lick plates have two exposures 
(I~=~long; II~=~short) and were taken with parallel-wire objective gratings.  
(See Section~IVa of Paper~I and Section~III of \citet{kvsw71} for details.)
Bright stars ($B \lesssim 14$) have many images, most of which are not useful
for astrometry, and many of which may overlap with other stars' images in the
dense NPM2 fields.  Besides extracting the $X,Y$ coordinates and $m_b$ or 
$m_y$ instrumental magnitudes, the exposure system (I or II) and grating order 
(0, 1, 2, or 3) of each image must be recognized.  

On the Lick plates, only those system/order combinations known (from NPM1) 
to give good astrometry were selected from the PMM scans.  For faint stars 
($B \gtrsim14$), only the long-exposure I\,(0) image is useful. For stars 
with $B \lesssim 14$, we use I\,(2), I\,(3), and II\,(0) on the blue plates; 
I\,(1), I\,(2), and II\,(0) on the yellow plates.  Grating images are always
taken as north-south pairs for orders 1, 2, or 3.  For the brightest 
stars ($B \lesssim 10$), only the short-exposure II\,(0) image is useful.
The task of selecting NPM2 stars from the PMM scans divided into three parts:

\begin{enumerate}

\item
For most NPM2 stars, image selection could be done automatically, i.e. without
first having to identify the stars at the Lick Survey Machine (as was done for
all stars in NPM1).  Stars with known, accurate positions, chiefly bright
($B \lesssim 13$) stars from Hipparcos, Tycho-2, and other astrometric
catalogs, could be identified once a transformation was established, for each
plate, between $\alpha,\delta$ and $X,Y$ on the PMM scans.  
We developed software, based partly on Yale SPM programs \citep{gir04}, 
to automate the process of extracting these stars' useful images (or image 
systems) from the PMM scans.  

\item
Faint anonymous astrometric reference stars (400 per field for NPM2), 
were also selected automatically.  We identified unblended I\,(0) images with
$ 15 \lesssim B \lesssim 17$, uniformly distributed over each field.
To ensure choosing the same star on all three plates (B1, B2, Y2), stars with
proper motion displacements (${\rm B2 - B1}$ or ${\rm Y2 - B1}$) larger than 
$\sim$30~$\mu$m were not selected.

\item
All NPM2 stars not already having accurate positions were identified at the 
Lick Survey Machine (often using finding charts).  Klemola surveyed all 347 
NPM2 fields to provide the $X,Y$ coordinates that allowed these stars' images 
to be selected from the PMM scans.  Most of the fainter ($B \gtrsim 12$)
``special stars'' chosen from the literature by Klemola for their astronomical
interest fell into this category.  Conservative procedures were followed to
avoid erroneous selections.  About 6\% of surveyed stars were not selected
from the B1 plate scans; another 4\% were ``lost'' on the B2 plate scans.

\end{enumerate}

For each NPM2 field, every star selected was indexed for cross-identification,
and the data from the PMM scans for each plate (B1, B2, Y2) were
organized into the same file structure used for NPM1, allowing the PMM data
to enter the NPM data reduction pipeline originally designed for AME plate
measurements.

\section{NPM2 Data Reductions} \label{RED}

Wide-field astrograph plates such as those of NPM and SPM are a unique
astrometric resource because they are amenable to the techniques of
differential astrometry necessary to determine accurate absolute proper
motions (Paper~I; Platais et~al.~1998a),
and can be measured with astrometric precision
over a magnitude range ($8 \lesssim B \lesssim 18$ for NPM; $5 \lesssim V
\lesssim 18$ for SPM) wide enough to encompass bright astrometric reference
stars {\it and}\/ faint galaxies.  Accurate astrometry is greatly facilitated
when plates of a field at multiple epochs can be closely ``matched'' (plate
center, hour angle, emulsion, exposure, etc.) in a way that can be achieved
with a dedicated, stable telescope such as the Lick astrograph.  
With well-matched plates and differential reductions, image irregularities
and magnitude-dependent effects have minimal impact on the proper motions.  
The NPM2 data reductions are designed to work differentially whenever possible,
to take fullest advantage of the Lick astrograph plates. 

To optimize accuracy, we have repeatedly tested and refined the procedures 
and plate models for determination of positions, proper motions, magnitudes,
and colors using the NPM plates.  For NPM2 we extensively revised the NPM1 
reduction programs (Sections~VI and VII of Paper~I), with more sophisticated 
plate models and rapid automated checking to delete blended images and 
erroneous measurements.  References to Paper~I will mark procedures that are 
still the same as for NPM1.  The structure of the NPM2 reductions is much the 
same as NPM1:  Each field is reduced independently.  Each step in the 
astrometric and photometric reductions of the B1, B2, and Y2 plates 
(or plate pairs; see below) produces plate constants, which are later used
to transform the $X,Y$ plate coordinates\footnote{The NPM2 reductions use
$X,Y$ coordinates in mm, with the origin at the plate center.  
$X,Y$ residuals will be expressed in \micron.}
and $m_b,m_y$ instrumental magnitudes into positions, proper motions, 
$B$~magnitudes, and \bv~colors.  To avoid spurious results, discordant or 
doubtful measures noted at any stage of the reductions are rejected.  
(Rejections average 5\% on blue plates, 3\% on yellow.)
Finally, multiple measures
of any star are checked and averaged to give one entry per star
(J2000 position at mean plate epoch $\sim$1968, absolute proper motion, 
$B$ magnitude, and \bv color).  The result is a set of 347 ``field catalogs'', 
which were compiled into the NPM2 Catalog as described in Section~\ref{CAT}.

\subsection{Reference Stars} \label{RFS}

NPM2 was able to use more sophisticated plate models than NPM1 because (1)~the
PMM data are full plate scans, supplying many more potential reference stars 
than NPM1, which was limited to the stars pre-selected for AME measurement;
and (2)~the release of the \mbox{Tycho-2} Catalogue in 2000 greatly increased 
the numbers of stars faint enough ($B > 8$) to be useful to NPM2 that have
accurate positions, proper motions, and photometry.  Using more reference stars
enabled major improvements in all three phases of the NPM2 reductions.  

The two-exposure NPM photography optimized image quality for ``bright'' stars 
($B \sim$12) and ``faint'' stars ($B \sim$16).  For the proper motion 
reductions, stars with $15 \lesssim B \lesssim 17$ are so abundant that 
we could take as many as needed (400 per field) from the PMM scans
(Section~\ref{STS}).
However, an ideal positional reference frame for the NPM plates at 
$11 \lesssim B \lesssim 13$ would extend at least 1~mag fainter than the
ACRS Catalog \citep{cor91}, which we used for NPM1.

Tycho-2 is nominally complete only to $V_T \sim 11$, but it actually contains 
nearly 1.7 million stars with $V_T > 11$ \citep[see Table 2 of][]{hog00b}, 
with moderately accurate positions ($60 < \sigma < 90$~mas) 
and proper motions ($2.5 < \sigma < 3$\masyr) in the ICRS/Hipparcos
system, and two-color ($B_T,V_T$) photometry ($0.1 < \sigma < 0.3$~mag).
Switching to Tycho-2 as our positional reference catalog allowed us to
use as many stars as necessary, in the ideal magnitude range, to model
the NPM2 plates for the highest possible positional accuracy.
From the PMM scans, we extracted 200 Tycho-2 stars distributed uniformly over 
each field, in the magnitude range $11 <B_T < 13$, in addition to a roughly 
equal number of (generally brighter) Tycho-2 stars already chosen for other 
reasons.

\subsection{Astrometric Plate Model} \label{PLT}

Having some 400 Tycho-2 reference stars allows the NPM2 positional reductions
to use the same cubic plate model as the proper motion reductions.  This 
desirable consistency had not been possible for NPM1, where having only 
$\sim$50 ACRS reference stars per field (see Table~\ref{tbl-1}) limited the 
positional plate model to a five-constant quadratic in each coordinate, 
and required pre-corrections for cubic radial distortion. Using the {\sl same}
astrometric model consistently for positions {\sl and} proper motions also has
the advantage that the two reductions can be compared, for example to check 
whether ($X,Y$)\,-\,dependent effects seen in the position reductions are also
evident in the proper motions (see below, and Section~\ref{PMMASK}).

The NPM2 astrometric plate model (Table~\ref{tbl-2}), optimized after extensive
test reductions, is a minor improvement on the 13-constant model we used for 
the NPM1 proper motion reductions (Section~VIIa of Paper~I).  These models 
represent a sophisticated compromise between (1)~a simple ``physical''
model with seven terms for zero point ($X,Y$), scale, rotation, plate tilt 
($P,Q$), and cubic radial distortion; and (2)~a full cubic model with separate 
terms in in each coordinate ($2 \times 10 = 20$ terms).  We can render several
of these terms unnecessary by pre-correcting the $X,Y$ measures
for differential aberration and refraction \citep[][Eqs.~20 and 25]{kon62}.

With a full cubic plate model, high correlations ($\rho > 0.9$) between 
the linear and cubic terms are always present.  NPM1 showed that these 
correlations can be substantially reduced (to $\sim 0.7$) by tying together 
the $X$ and $Y$ solutions by a common scale coefficient, which is possible 
when differential refraction is pre-corrected.  Moreover, several terms 
($Y^2$ and $Y^3$ in $X$; $X^2$ and $X^3$ in $Y$) with no clear physical 
significance proved to be negligible in test reductions, and could be 
eliminated from the model.

With 400 reference stars per NPM2 field, we can determine a full set of cubic 
distortion terms ($X^3$ and $XY^2$ in $X$; $X^2Y$ and $Y^3$ in $Y$) on each 
plate (positions) or plate pair (proper motions) instead of the fixed
radial-distortion pre-corrections NPM1 used.  Finally, we now solve for the 
diagonal distortion-like cubic term ($X^2Y$ in $X$; $XY^2$ in $Y$) discovered 
in NPM1.  

The NPM2 plate model (Table~\ref{tbl-2}) has 14 terms, one more than the 
NPM1 proper motion model.  The terms in the NPM2 model that would be equal in
a simple seven-constant physical model are those for rotation ($c_2=c_3$), 
plate tilt ($c_6=c_8=P$, $c_7=c_9=Q$), and radial distortion 
($c_{10}=c_{11}=c_{12}=c_{13}$).  In the NPM2 reductions, these terms 
usually show small but significant differences, because they absorb unmodeled 
effects such as higher-order or radially-asymmetric distortion. 

For the most part, the numerical values of the position and proper motion
plate constants have little scientific importance, tending simply to reflect
the circumstances of observation and measurement.  However, three points 
regarding the cubic terms are worth noting: (1)~The NPM2 distortion constants 
are quite small (mean size $\lesssim 10^{-9}$~mm$^{-2}$, amounting to 
10--20~\micron\ at the plate edges and corners), but highly significant 
(usually 5--10~$\sigma$).  This is like what \citet{plx95} found for the 
SPM plates, but we find no significant trends of the NPM2 constants over time.
(2)~Mean values of the NPM2 constants nearly equal the values adopted in the 
NPM1 position reductions or determined in the NPM1 proper motion reductions.
(3)~Field by field, the {\sl differences} between the NPM2 B1 and B2 position 
constants roughly equal the the corresponding proper motion plate constants
determined by differential reduction of the B1 and B2 plates.  

We note that the NPM2 plate model does {\sl not} include magnitude-dependent 
terms.  For positions, the magnitude range ($9 < B < 13$) of the Tycho-2 
reference stars is not large enough to determine such terms reliably on each 
plate (B1, B2, Y2).  Moreover, there is no reason to expect that the {\it same}
terms would apply to the faint stars ($14 < B < 18$) that form the bulk of 
the NPM program.  The possibility of magnitude-dependent corrections to the 
NPM2 positions is considered in Section~\ref{MAG}.  For proper motions, we use 
faint anonymous stars ($ 15 \lesssim B \lesssim 17$) on each (B1,\,B2) plate 
pair.  Magnitude terms cannot be used in the plate solutions because such terms
would absorb real effects (e.g., secular parallaxes) in the stellar motions.
Magnitude-dependent systematic proper motion errors for our bright stars 
$(8 < B < 12)$ are corrected in each field as part of the final correction 
from relative to absolute proper motions, discussed in Section~\ref{PMABS}.

\section{Positional Reductions} \label{POS}

For each of the 347 NPM2 fields, the reduction of the PMM $X,Y$ measures 
to J2000 positions $\alpha, \delta$ is done in four stages.  
Until the final stage, all three plates (B1, B2, and Y2) are reduced 
separately, as follows.

\subsection{Bridge Reductions and Pre-Corrections} \label{BRG}

As in NPM1 (Section~VIa of Paper~I), the first step in the NPM2 plate
reductions is the positional ``bridge reduction'', which differentially
reduces bright stars' short-exposure images into the $X,Y$ system of the 
long exposure.  First, on each plate, all (north-south) diffraction-grating 
image pairs I\,(gr) are averaged.  The average is equivalent to the 
I\,(0) position\footnote{The SPM \citep{gir98,gir04} procedure, which 
explicitly reduces grating images to the zero-order system, cannot be used 
on the NPM plates, where the I\,(gr) and I\,(0) images are never useful 
for the same star.}, neglecting small magnitude- and color-dependent effects 
(see Section~\ref{MAG}).
Then, all stars having I\,(gr) {\sl and} II\,(0) images (usually over 200 
stars on B1 and B2, near 400 on Y2; 4--6 times as many as NPM1) are identified.
As in NPM1, the II\,(0) images are corrected for differential plate tilt, so 
that a simple four-constant solution (scale, rotation, and $X,Y$ zero point)
can transform all II\,(0) $X,Y$ measures on the plate to the I\,(0) system.
The unit weight errors of the bridge solutions are generally 3--4~\micron\
on B1 and B2, and 2~\micron\ on Y2.  These are about twice the PMM $X,Y$ 
errors, largely because many poor (e.g. weak, overexposed, or blended) images
have not yet been rejected.)  The contribution of the bridge transformation
to the positional error for a bright star's II\,(0) image is very small,
about 12--16~mas on B1 and B2, and 6~mas on Y2.

Next, multiple measures of one star are checked for consistency of position.
Stars with discordant $X,Y$ measures (more than 5~$\mu$m from the mean)
are examined to identify which image system should be rejected.  If that
fails (e.g., when there are only two measures) the star is totally rejected.
Generally, 1-2\% of all images are rejected here; most of these are blended
images not recognized at the PMM star selection stage.

After the bridge reduction, all the $X,Y$ measures on the plate, for bright 
and faint stars alike, are on the I\,(0) system of the faint stars.  
Finally, we pre-correct $X,Y$ for differential aberration and refraction,
as in NPM1.
The reduced $X,Y$ values for each plate (B1, B2, Y2) are then used as input 
to the plate-constant reductions for positions and proper motions.  

\subsection{Plate Constant Reductions} \label{PPCR}

All positional reference stars used for NPM2 come from the Tycho-2 Catalogue,
which lists mean positions and proper motions in the ICRS system.  
Tycho-2 positions are given for the catalog epoch J2000, with the Tycho-2 
proper motions having been used to propagate the positions from the mean 
observational epoch $T_0$ (averaging $\sim$1980 but ranging from $\sim$1912 
to $\sim$1992) to the J2000 epoch.  Position errors $\sigma_0$ 
in each coordinate are given at the mean epoch $T_0$, where position and 
proper motion are uncorrelated, and the positional error is minimized.  
These data are sufficient to propagate the Tycho-2 positions and their errors 
to any other epoch.

For each NPM2 plate (B1, B2, Y2), we backdate the Tycho-2 positions (J2000 
coordinates and epoch) to the Lick plate epoch $T_1$ (near 1950 for B1; 
usually 1970--1980 for B2,Y2) using the Tycho-2 proper motions.  In each 
coordinate the Tycho-2 position error $\sigma_1$ at the Lick plate epoch is 
the quadratic sum of the pure positional error ($\sigma_0$) and the proper 
motion error ($\sigma_\mu$) accumulated over $\vert T_1 - T_0\vert$ years:
\begin{equation}
\sigma_1^2\ \ =\ \ \sigma_0^2\ +\ \sigma_\mu^2\,(T_1 - T_0)^2.
\end{equation}
The errors $\sigma_1$ of the backdated Tycho-2 positions used for NPM2 
average $\sim$90~mas (1.6~$\mu$m on the NPM plates) at the Lick first epoch,
and $\sim$60~mas (1.1~$\mu$m) at the second epoch.  This is roughly the size
of the PMM measurement errors. 

On each plate, the backdated Tycho-2 positions are used to compute standard
coordinates $X_T, Y_T$ (in millimeters from the plate center) for each star.
The PMM measures for each Tycho-2 star are averaged over multiple image systems
if necessary.  The measured coordinates $X_M,Y_M$ are then adjusted to the
standard coordinates in a least-squares solution with the equations of
condition

\begin{mathletters}
\begin{equation}
\Delta X\ =\ X_T - X_M\ =\ \sum_{i=1}^{14}\, c_i\, T_{i,X} 
\end{equation}
and
\begin{equation}
\Delta Y\ =\ Y_T - Y_M\ =\ \sum_{i=1}^{14}\, c_i\, T_{i,Y}
\end{equation}
\end{mathletters}

\noindent
where $c_i$ are the 14 plate constants and $T_{i,X},\,T_{i,Y}$
are the corresponding $X,Y$ terms in Table~\ref{tbl-2},
evaluated using $X_M,Y_M$.
Terms not used in $X$ have $T_{i,X} =0$; likewise for $Y$.
We define the $\Delta X$ and $\Delta Y$ differences
as ``catalog minus measured'', to be consistent with NPM1.
Weights are determined from the Tycho-2 errors (Eq.~1) and nominal values 
of the PMM measurement errors.  Each star's residuals are

\begin{mathletters}
\begin{equation}
R_X\ \ =\ \ X_T - X_S\ \ =\ \ X_T\, -\, (X_M+\Delta X),
\end{equation}
and
\begin{equation}
R_Y\ \ =\ \ Y_T - Y_S\ \ =\ \ Y_T\, -\, (Y_M+\Delta Y),
\end{equation}
\end{mathletters}

\noindent
where $X_S,Y_S$ are the standard coordinates corresponding to $X_M,Y_M$.
Stars with large residuals are rejected and the solutions repeated.
After convergence, the numbers of Tycho-2 stars surviving average 350 per 
blue plate, and 390 per yellow plate.

Over all 347 NPM2 fields, the RMS unit weight errors $\sigma_{pos}$ of the 
position solutions are 3.5~$\mu$m (190~mas) for B1 plates, 2.9~$\mu$m (160~mas)
for B2, and 1.8~$\mu$m (100~mas) for Y2.  The differences chiefly reflect that 
the Lick yellow plates have sharper images, and the Tycho-2 positions are 
generally more accurate near NPM2's second epoch.  

\subsection{Positional Plate Masks} \label{POSMSK}

Because the NPM plate model cannot be a perfect representation of the Lick 
plates, the position residuals will partly reflect whatever unmodeled 
(e.g., higher-order) positional effects are actually present on the plates.
We can check for, and largely correct, these effects by the method of 
``stacking'' the $347 \times 3$ sets of residuals $R_X,R_Y$ to make 
``plate masks.''  We stress the importance of doing this separately for the 
B1, B2, and Y2 plates.

\begin{figure}
\includegraphics[scale=0.665]{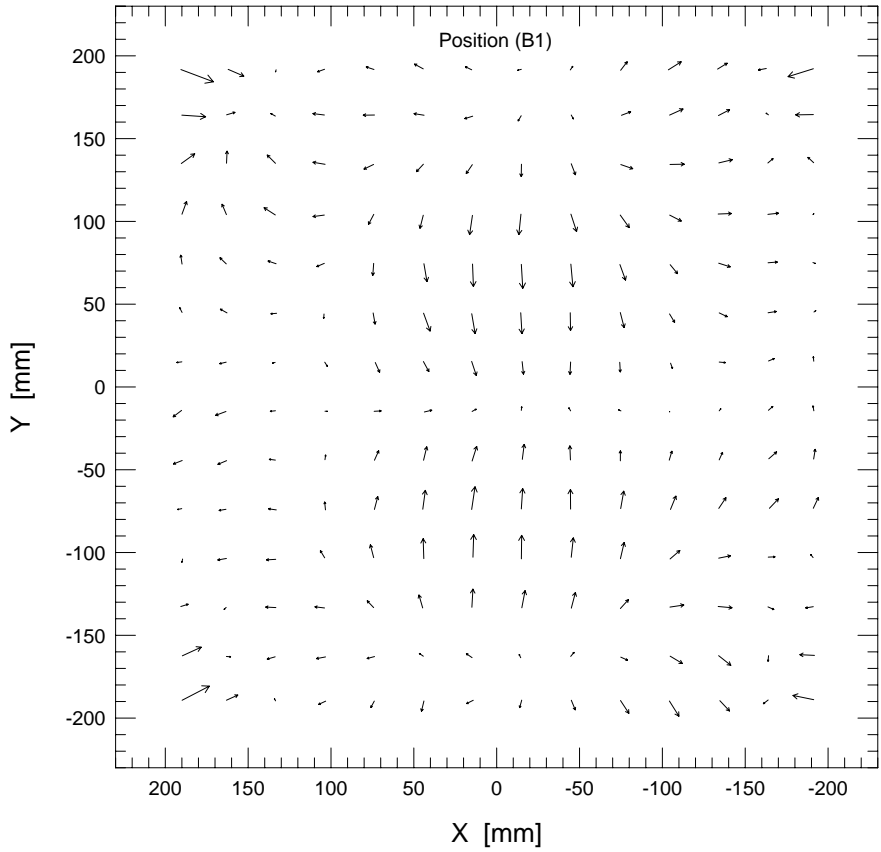}
\caption[fig2.eps]{
Position ``mask'' for NPM2 first epoch blue plates (B1).
$X,Y$ are Lick plate coordinates.
Vectors show the mean residuals $\langle R_X \rangle$, $\langle R_Y \rangle$ 
for 122,759 Tycho-2 stars in $14 \times 14 = 196$ bins 
(average 626 stars per bin).
Vector length scale is 10~mm = 1~$\mu$m.
RMS vector length is 0.75~$\mu$m (42~mas).
Longest vector is 2.10~$\mu$m (116~mas).
\label{pgB1}
}
\end{figure}

\begin{figure}
\includegraphics[scale=0.665]{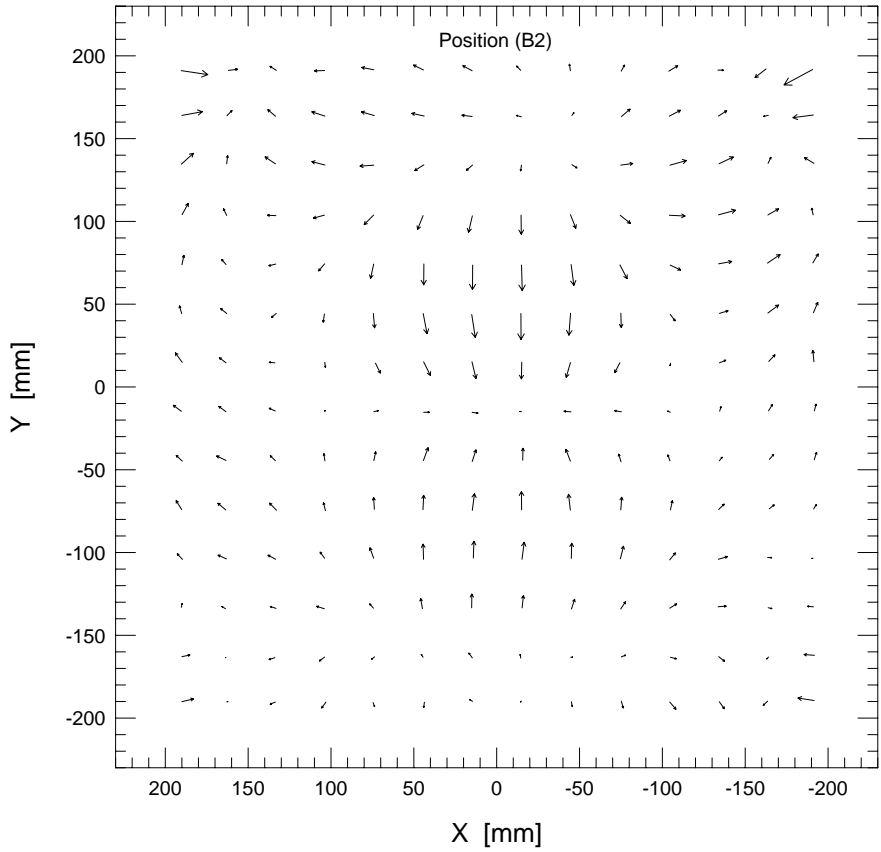}
\caption[fig3.eps]{
Position ``mask'' for NPM2 second epoch blue plates (B2).
$X,Y$ are Lick plate coordinates.
Vectors show the mean residuals $\langle R_X \rangle$, $\langle R_Y \rangle$ 
for 118,349 Tycho-2 stars in $14 \times 14 = 196$ bins 
(average 604 stars per bin).
Vector length scale is 10~mm = 1~$\mu$m.
RMS vector length is 0.69~$\mu$m (38~mas).
Longest vector is 1.95~$\mu$m (108~mas).
\label{pgB2}
}
\end{figure}

\begin{figure}
\includegraphics[scale=0.665]{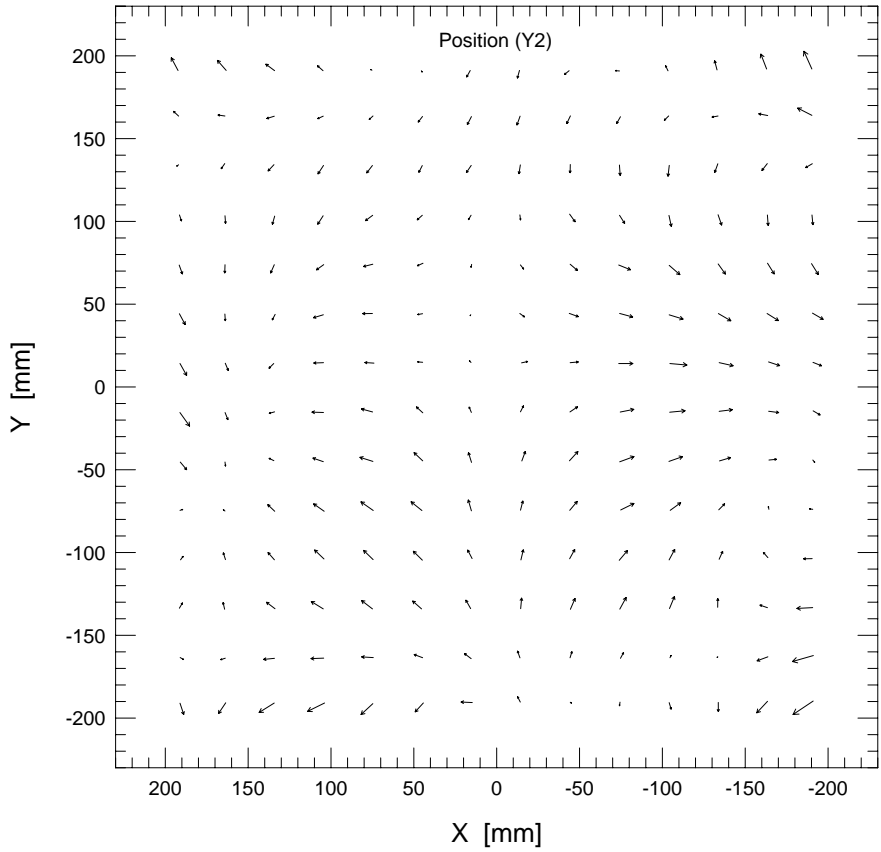}
\caption[fig4.eps]{
Position ``mask'' for NPM2 second epoch yellow plates (Y2).
$X,Y$ are Lick plate coordinates.
Vectors show the mean residuals $\langle R_X \rangle$, $\langle R_Y \rangle$ 
for 133,516 Tycho-2 stars in $14 \times 14 = 196$ bins 
(average 681 stars per bin).
Vector length scale is 10~mm = 1~$\mu$m.
RMS vector length is 0.62~$\mu$m (34~mas).
Longest vector is 1.46~$\mu$m (80~mas).
\label{pgY2}
}
\end{figure}

\begin{figure}
\includegraphics[scale=0.665]{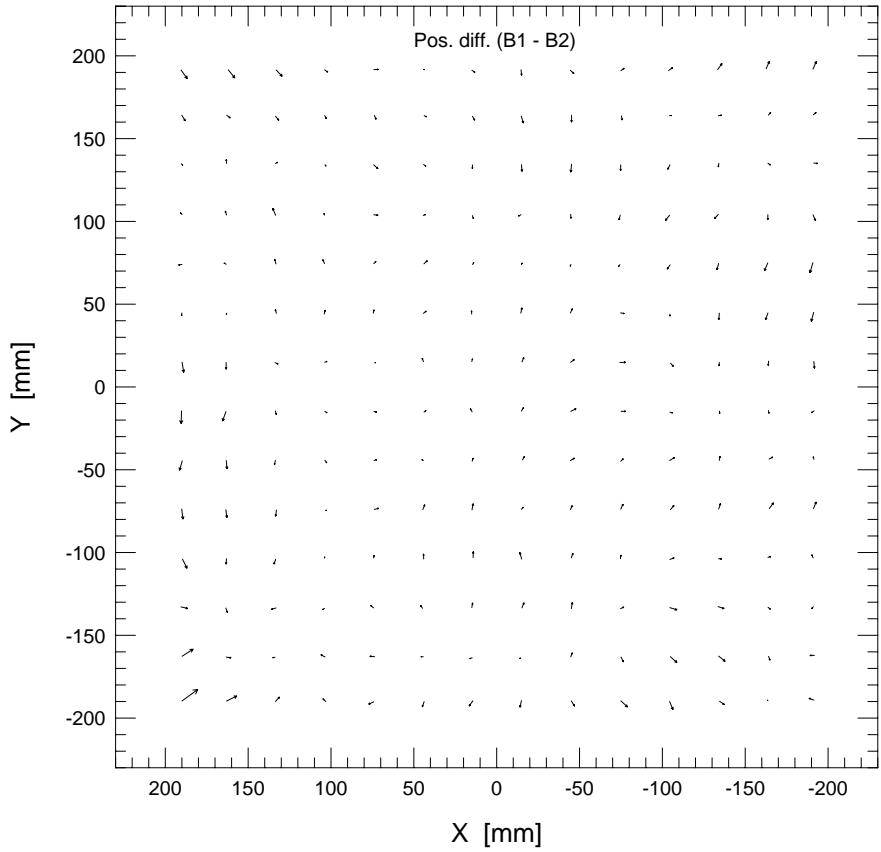}
\caption[fig5.eps]{
Difference ``mask'' (B1 $-$ B2) for NPM2 positions.
$X,Y$ are Lick plate coordinates.
Vectors show the differences between the B1 and B2 masks 
(Figs.~\ref{pgB1} and \ref{pgB2}) in $14 \times 14 = 196$ bins.
Vector length scale is 10~mm = 1~$\mu$m.
RMS vector length is 0.33~$\mu$m (18~mas).
Longest vector is 1.17~$\mu$m (65~mas).
See text for comparison with Figure~\ref{pmgrid}.
\label{pdB12}
}
\end{figure}

Figures~\ref{pgB1}, \ref{pgB2}, and \ref{pgY2} are the positional plate masks
for the B1, B2, and Y2 plates, respectively.  We divide the plate into
$14 \times 14 = 196\ \ X,Y$ bins, because the faint half of the Tycho-2 stars
were selected in such bins to enforce a uniform distribution of reference
stars.  Each vector is centered at the mean $X,Y$ position for the bin,
and typically represents the mean residuals $\langle R_X \rangle$, 
$\langle R_Y \rangle$ of $\langle N_{bin} \rangle \simeq 600$ stars.
Several features of the NPM2 plate masks are worth discussing in detail:

\begin{enumerate}

\item
All three masks (B1, B2, Y2) reveal small (RMS~$\sim 0.7~\mu$m $\simeq 40$~mas)
but highly significant ($4\,\sigma$ for B1 and B2; $6\,\sigma$ for Y2) 
systematic patterns that could not have been modeled by cubic terms.  

\item
The blue plate patterns for the two epochs (Figures~\ref{pgB1} and \ref{pgB2})
are remarkably similar, a testament to the long-term stability 
of the Lick astrograph, which is vital to the determination of accurate 
proper motions. Figure~\ref{pdB12} shows the B1 $-$ B2 pattern differences;
the RMS vector difference is 0.4~$\mu$m $\simeq 20$~mas. (See Section~\ref{PMS}
for a comparison with the plate mask from the NPM2 proper motion reductions.)

\item
The yellow plate pattern (Figure~\ref{pgY2}) is quite different from the blue,
even though the B2 and Y2 plates for each field were exposed simultaneously.
This suggests that the plate patterns reflect optical characteristics of the 
astrograph's two lenses, rather than observational effects.

\item
The NPM position masks are rather different from the
SPM astrograph's position mask \citep[Figure 4b of][]{gir04}, which clearly
reveals fifth-order radial distortion effects not seen on the NPM plates.  

\item
The NPM Y2 plate pattern (Figure~\ref{pgY2}) is radically different from 
Figure~9 of \citet{zac04}, which illustrates the ``field distortion pattern''
(FDP) of NPM Y2 plates vs. Tycho-2, as found in plate reductions for the 
(unpublished) USNO ``Yellow Sky 3'' (YS3) catalog .  Apparently, that FDP
is an artifact of the YS3 reductions.

\end{enumerate}

After we use the plate masks to correct the $X,Y$ positions on each plate
(B1, B2, Y2), any remaining field-dependent errors are reduced to the noise 
level set by\, $\sigma_{pos}\, /\, \surd\langle N_{bin} \rangle$, 
which amounts to $(0.14,\, 0.12,\, 0.07)~\mu$m, or $(7.7,\, 6.5,\, 3.8)$~mas, 
respectively.

\subsection{Combined Positions} \label{POSC}

For each NPM2 field, the final stage in the positional reductions applies the 
plate constants and the plate-mask corrections to the $X,Y$ measures, 
transforms $X,Y$ to $\alpha, \delta$, averages the three positions 
(B1, B2, Y2) for each star, and checks for discordances.

On each plate, all multiple measures for a given star are averaged.  
Then, for each star, the 14 positional plate constants (Eqs.~2a and 2b) 
are applied to transform $X_M,Y_M$ into the system of the standard coordinates 
$X_S = X_M + \Delta X,\ Y_S = Y_M + \Delta Y$.  Next, the appropriate 
plate-mask correction (B1, B2, or Y2) is applied by adding the mean residuals 
 
\begin{mathletters}
\begin{equation}
X_C\ \ =\ \ X_S\, +\, \langle R_X \rangle\ =\ \ X_S\, 
                  +\, \langle X_T - X_S\rangle
\end{equation}
and
\begin{equation}
Y_C\ \ =\ \ Y_S\, +\, \langle R_Y \rangle\ =\ \ Y_S\, 
                  +\, \langle Y_T - Y_S\rangle
\end{equation}
\end{mathletters}

\noindent
where $\langle R_X \rangle$ and  $\langle R_Y \rangle$ are interpolated in the 
14$\times$14 $X,Y$ grid.  The mask-corrected standard coordinates $X_C,Y_C$, 
now in the Tycho-2 system, are then transformed into J2000 $\alpha, \delta$.

The result of the plate reductions is three (two if there is no Y2 measure) 
separate positions for each star, each at the plate epoch.  These are averaged
with equal weight, as was done in NPM1.  To check for positional discrepancies,
we apply each star's NPM2 proper motion (as determined in Section~\ref{PMS}) 
to bring each position (B1, B2, Y2) from its plate epoch to the mean epoch.  
The difference between each position and the mean position is calculated.  
Any star having a residual larger than $0\,\farcs5$ ($\sim$9~$\mu$m) 
in either coordinate is identified with an error code in the ``field catalog'' 
for that field.  Any star with a residual larger than $3\,\farcs0$ 
($\sim$54~$\mu$m) is rejected.
(The rejection rate at this stage is only $\sim$0.1\%.)

The internal errors of the NPM2 mean positions, {\sl at the mean epoch of 
observation,} $\sim$1968, average $\sim$80~mas (RMS) in each coordinate.  
These results represent a factor-of-two improvement over the precision obtained
in NPM1 using a simple quadratic plate model and ACRS reference stars.  
As noted in (Section~\ref{POSMSK}, field-dependent systematic errors of the 
NPM2 positions are well below 10~mas.  This is nearly an order of magnitude 
better than NPM1, thanks to the Tycho-2-based reductions and the plate-mask 
corrections.

\subsection{Magnitude-Dependent Effects} \label{MAG}

From our experience with NPM1, we expect that magnitude-dependent effects, 
on the order of 1-2~$\mu$m in size, may be present in the NPM2 positions.
Two questions need to be considered.  First, are there ``magnitude equations''
due to the lack of magnitude terms in the NPM2 plate model 
(Section~\ref{PLT})?  Second, how well does the bridge reduction 
(Section~\ref{BRG}) put the bright and faint NPM2 stars on the same system?

Effects that are very small in the context of the individual plate reductions,
with several hundred stars per plate, can be detected by comparison of NPM2 
with other large positional catalogs, such as Tycho-2 or the new UCAC2 Catalog
\citep{zac03,zac04}.  

For bright stars ($B < 13$), we used some 370,000 residuals
(Tycho-2 $-$ NPM2) from the B1, B2, and Y2 plate solutions.  
These were examined versus B~magnitude and $B - V$ color.
Small non-linear magnitude effects are seen, with slopes 
up to 1~$\mu$m~mag$^{-1}$ (55~mas~mag$^{-1}$) in $X$,
and 0.5~$\mu$m~mag$^{-1}$ (30~mas~mag$^{-1}$) in $Y$.
These results are very similar to those previously found by N.~Zacharias 
(2003, private communication), who compared the NPM2 Catalog positions 
with the UCAC2 Catalog.
In addition, the Tycho-2 $-$ NPM2 comparison indicates that these effects 
may also vary with $B-V$, declination, and plate epoch (B1~vs.~B2) 
or plate color (B~vs.~Y). 
For NPM2 faint stars ($13 < m_R < 16$; roughly 
$14 < B < 17$), Zacharias found linear magnitude effects up to
40~mas~mag$^{-1}$ (0.7~$\mu$m~mag$^{-1}$) in size.

We have chosen not to correct the NPM2 positions for these magnitude-dependent
effects, which are small in the context of NPM2's positional precision
(80~mas near 1968, 200~mas at 2000), and problematical to calibrate in terms 
of all the observational variables.  Users are reminded that NPM is principally
a proper motion program.  To the extent that these positional effects may vary
between epochs, they would affect the NPM2 proper motions, so we do apply
magnitude-dependent proper motion corrections (Section~\ref{PMABS}).

The NPM2~$-$~UCAC2 comparison (Zacharias~2003) also verifies that the NPM2 
bridge reductions accurately put our bright stars (measured as I\,(gr) and 
II\,(0) images) and faint stars (measured as I\,(0) images) onto the same 
positional system.  In both right ascension and declination, stars with
$m_R \sim 11\ \ (B \sim 12$) and 
$m_R \sim 15\ \ (B \sim 16$) systematically agree in position
to within 10 to 20~mas (0.2 to 0.4~\micron).

\section{Proper Motions} \label{PMS}

Absolute proper motions\footnote{
In the NPM program we follow the usual convention of focal-plane astrometry,
defining $\mu_\alpha = \mu\sin\theta$ and $\mu_\delta = \mu\cos\theta$, 
where $\mu$ is the total proper motion, and $\theta$ is its position angle
in local celestial coordinates (north = 0\degr, east = 90\degr).  
Thus, both proper motion components are expressed in ``great circle measure'',
in the same units, ready to use for stellar kinematics. 
By comparison, positional astronomy usually defines
$\mu_\alpha =\ $d$\alpha/$dt, in terms of an angle subtended at the Pole,
which is the form needed to update stellar positions.
In Hipparcos' notation \citep[][Vol.~1, Sec.~1.2.5]{esa97},
${\rm \ \mu_{\alpha*} = (d\alpha/dt) \cos\delta }$, which is equivalent to
NPM's $\mu_\alpha$.
}
($\mu_\alpha,\,\mu_\delta$) in each of the 347 NPM2 fields were determined 
in three steps.  First, we perform a differential plate-constant reduction 
of the B1,\,B2 plate pair, to obtain relative proper motions.  (As in NPM1, 
the yellow plate Y2 is not used for proper motions, because there is no Y1
plate for differential reductions.) Second, we apply a plate-mask correction 
for ($X,Y$)-dependent systematics. Third, we correct magnitude-dependent 
systematic errors and reduce the relative proper motions to absolute.
This Section will describe these steps, and assess the accuracy of the
NPM2 absolute proper motions.

\subsection{Plate Constant Reductions} \label{PMSC}

As in NPM1, we determine relative proper motions by a differential 
plate-constant reduction of the blue plates at the two epochs.
We use about 400 faint anonymous stars (``FAS''; Section~\ref{STS}) per field, 
mostly in the range $15 < B < 17$, as~astrometric reference stars.  
As discussed in Section~\ref{PLT}, we use the same plate model
(Table~\ref{tbl-2}) as in the positional reductions.  
The B1 plate coordinates $X_1,Y_1$\/ are adjusted to the B2 coordinates 
$X_2,Y_2$\/ in a least-squares solution with the equations of condition

\begin{mathletters}
\begin{equation}
\Delta X\ =\ X_2 - X_1\ =\ \sum_{i=1}^{14}\, c_i\, T_{i,X} 
\end{equation}
and
\begin{equation}
\Delta Y\ =\ Y_2 - Y_1\ =\ \sum_{i=1}^{14}\, c_i\, T_{i,Y}
\end{equation}
\end{mathletters}

\noindent
As before, $c_i$ are the 14 plate constants.  The terms
$T_{i,X},\,T_{i,Y}$ are evaluated using $X_1,Y_1$.  
All equations are weighted equally, because most 
of the variance is ``cosmic'', i.e. the residuals 
\begin{equation}
R_X\ =\ X_2\, -\, (X_1 + \Delta X)\ \ \ \ {\rm and}\ \ \ \
R_Y\ =\ Y_2\, -\, (Y_1 + \Delta Y)
\end{equation}
are largely due to the stellar proper motions.  Because the distribution of 
proper motions in a small area of the sky \citep[][Sec.~3.52]{tw53} generally 
has a sharp core and asymmetric wings, severe trimming of large residuals
is necessary.  Any star with a vector residual (proper motion) larger than 
20~$\mu$m (1.1~arcsec) in the first iteration is rejected.  Subsequent 
iterations reject $3\,\sigma$ residuals.  After convergence, the average 
number of FAS remaining is about 350.  Finally, any FAS with a proper motion
larger than 30~$\mu$m is flagged for rejection from the NPM2 Catalog.  
Tests showed that such large FAS motions are occasional spurious results 
(average 2 per field) of the automated FAS selection process.

Over all 347 NPM2 fields, the RMS unit weight errors of the proper motion 
plate constant solutions are $\sigma_{pm} = 4.0~\micron$ (220~mas).  
This corresponds to a proper motion dispersion of 8\masyr\ at the average 
epoch difference $\langle \Delta T \rangle = 28$~yr.  
It is important to note that $\sigma_{pm}$ is {\sl not} an error estimate 
for the NPM2 proper motions,  whose PMM measurement precision is
$\sim 2~\micron\ \times \sqrt2\ =\ 2.8~\micron$ (160~mas), corresponding to 
5.6~\masyr; and whose external accuracy $\sigma_\mu$ will be evaluated in 
Sections~\ref{PMACC} and \ref{ERR}.

\subsection{Proper Motion Plate Mask} \label{PMMASK}

\begin{figure}
\includegraphics[scale=0.645]{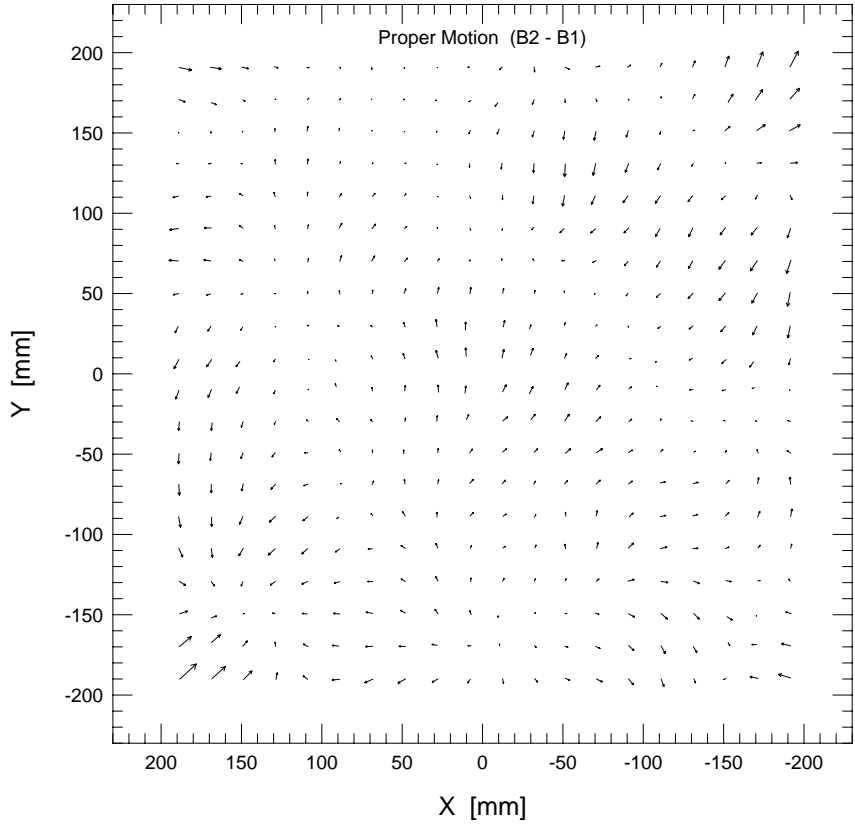}
\caption[fig6.eps]{
``Mask'' for NPM2 relative proper motions (B2 $-$ B1).
$X,Y$ are Lick plate coordinates.
Vectors show the mean proper motion displacements 
$\langle \Delta X \rangle$, $\langle \Delta Y \rangle$ 
for 119,954 faint anonymous stars in $20 \times 20 = 400$ bins 
(average 300 stars per bin).
Vector length scale is 10~mm = 1~$\mu$m.
RMS vector length is 0.38~$\mu$m (21~mas).
Longest vector is 1.39~$\mu$m (77~mas).
See text for comparison with Figure~\ref{pdB12}.
\label{pmgrid}
}
\end{figure}

Like the position residuals (Section~\ref{POSMSK}), the proper motion residuals
$R_X,R_Y$ will partly reflect whatever systematic effects are present on the 
plates but not in the model.  
Again, we can check and correct these effects by stacking the $X,Y$ proper 
motion residuals to make the mask shown in Figure~\ref{pmgrid}.  
Here we divide the plate into $20 \times 20 = 400\ \ X,Y$\/ bins, because the 
FAS were selected in such bins to enforce a uniform distribution on the plate.
Each vector represents the mean residuals $\langle R_X \rangle$, 
$\langle R_Y \rangle$ of an average of 300 stars.
The B2 $-$ B1 proper motion mask shows small (RMS~$\sim 0.4~\mu$m $\simeq 
20$~mas), complex systematic patterns that are marginally significant 
(1-2\,$\sigma$) for the individual bins, whose precision is limited to
$\sigma_{pm}\,/\,\surd 300\ \simeq\ 0.2~\mu$m.  To reduce the scatter,
Figure~\ref{pmgrid} shows a $3 \times 3$ smoothing of the mean residuals.
We use this mask to correct the proper motions of all stars in each NPM2 
field.

The plate mask from the NPM2 differential proper motion reductions must
reflect changes in the positional patterns of the Lick astrograph's blue lens 
between the two epochs, so Figure~\ref{pmgrid} should resemble the B1 $-$ B2 
positional difference\footnote{
The sign of the differences is reversed here because Eq.~2 defines the 
position residuals as ``catalog minus measured.''}
pattern in Figure~\ref{pdB12}.
Indeed, comparison of the two masks shows strong similarities in the size and
configuration of the patterns, especially near the plate corners and edges
where the effects are the largest.  
The similarity is particularly remarkable because of the totally different 
sets of stars and methods of reduction the two masks represent.  
The proper motion reductions use the long-exposure I\,(0) images of faint 
stars ($15 < B < 17$) and a differential reduction.  The position reductions 
use the short-exposure II\,(0) and long-exposure I\,(gr) images, 
and an absolute reduction of each plate.  These results give confidence
that our astrometric plate model is sound and has been successfully applied to
the wide range of magnitudes of the stars in NPM2.

\subsection{Relative Proper Motions} \label{PMREL}

After the FAS plate-constant reductions for a field are done, we calculate the
relative proper-motions for all stars, as follows.  First, multiple measures 
for a given star are averaged.  Then, for each star, the 14 plate constants
(Eqs.~5a and 5b) are applied to transform its measured B1 plate coordinates 
$X_1,Y_1$\/ into the $X_2,Y_2$\/ system of the B2 plate, i.e.
\begin{equation}
X_{12} = X_1 + \Delta X\ \ \ \ {\rm and}\ \ \ \ Y_{12} = Y_1 + \Delta Y.
\end{equation}
The differences (in the sense B2 $-$ B1) 
are the raw proper motion displacements
\begin{equation}
D_X = X_2 - X_{12}\ \ \ \ {\rm and}\ \ \ \ D_Y = Y_2 - Y_{12}
\end{equation}
to which the mask correction is applied by subtracting the 
faint reference stars' mean residuals, so that the corrected 
proper motion displacements $P_X,P_Y$ are
\begin{equation}
P_X = D_X - \langle R_X \rangle\ \ \ \ {\rm and}\ \ \ \
P_Y = D_Y - \langle R_Y \rangle
\end{equation}
where $\langle R_X \rangle$ and $\langle R_Y \rangle$ are interpolated in the 
20$\times$20 $X,Y$ grid.  Next, each star's relative proper motion components
are
\begin{equation}
\mu_X = S_B \times P_X\, / \, \Delta T\ \ \ \ {\rm and}\ \ \ \
\mu_Y = S_B \times P_Y\, / \, \Delta T
\end{equation}
where $\Delta T$ is the epoch difference (B2 $-$ B1) for that NPM field, 
and $S_B = 55.142$ arcsec~mm$^{-1}$ (or mas~\micron$^{-1}$) is the nominal
scale of the astrograph's blue lens.

Finally, we transform the relative proper motion from plate coordinates 
$\mu_X,\mu_Y$\/ into the equatorial components $\mu_\alpha,\mu_\delta$.  
On the $6\degr \times 6\degr$ NPM plates, this is approximately the simple 
rotation used in NPM1 \citep[][Fig.~4]{kvsw71} to orient $\mu_\delta$ 
toward the North Celestial Pole.  However, on the gnomonic plate projection,
$\mu_\alpha$ and $\mu_\delta$ are not strictly orthogonal, and the scale
varies slightly over the plate.
For NPM2 we use an exact geometric transformation, based on Tissot's 
indicatrix (the ``ellipse of distortion''), 
well-known in cartography \citep{mal92,sny97,can02} but not in astrometry.

\subsection{Corrections to Absolute} \label{PMABS}

Relative proper motions are not useful until they are corrected to an absolute
zero point by adding the actual mean motions of the reference stars, determined
by other means.  NPM1 corrected its proper motions to an absolute system, 
defined in each field by the mean motion of faint ($16 < B < 18$) galaxies 
relative to the FAS reference frame.  The accuracy of these corrections 
averaged 2\masyr\ (RMS, per field).  Magnitude-dependent corrections were not 
used, but tests showed that for $B > 12$, the bright and faint stars are 
on the same system, to within 1--2\masyr.  For the brightest NPM1 stars 
($8 < B < 12$), magnitude-dependent systematic errors 
\citep[``magnitude equations'';][]{plx98b} on the order of 1\masyr~mag$^{-1}$ 
in $\mu_\alpha$ and $\mu_\delta$ were found by comparing NPM1 with Hipparcos.

NPM2 has no reference galaxies, so a different method must be used.  For NPM2,
we correct the relative proper motions to the system of the Tycho-2 Catalogue, 
as explained below.  In principle this procedure is equivalent to a 
correction to absolute proper motion, because the Tycho-2 proper motions 
\citep{hog00a} are on the ICRS/Hipparcos proper motion frame \citep{kov97}, 
which was linked (rotated) to the inertial extragalactic reference frame 
defined by NPM1, SPM, and other absolute proper motions \citep{plx98b}.  
In practice, the accuracy of this equivalence fully satisfies the needs 
of NPM2 (see Section~\ref{PMACC}).

A great advantage of correcting NPM2 to Tycho-2 is that we can calibrate
and correct the NPM2 bright stars' magnitude equations in each coordinate 
as an integral part of the process.  We use Tycho-2, rather than Hipparcos
as we had originally intended \citep{han97}, because we now have some 350
Tycho-2 stars per field in the needed magnitude range ($9 < B < 13$).
This is ten times the number of Hipparcos stars useful for this purpose. 
Tests showed the precision of the NPM2--Tycho-2 corrections in an average 
field is $\sim$0.5\masyr, a factor of three better than we could have achieved
using Hipparcos stars.  

The magnitude equations vary substantially from field to field, so we 
calibrate and correct them separately in each NPM2 field.  To do this, 
we used  $\Delta\mu_\alpha$ and $\Delta\mu_\delta$\/ (Tycho-2 absolute minus 
NPM2 relative), with a least-squares solution in each coordinate, with
the equations of condition

\begin{mathletters}
\begin{equation}
\Delta\mu_\alpha\ =\ a_0\ +\ a_1\,(B-12)\ +\ a_2\,(B-12)^2
\end{equation}
and
\begin{equation}
\Delta\mu_\delta\ =\ b_0\ +\ b_1\,(B-12)\ +\ b_2\,(B-12)^2
\end{equation}
\end{mathletters}
\noindent

The B magnitudes are the NPM2 photometry, discussed in Section~\ref{PHOT}. 
The comparison is restricted to the magnitude range $9 < B < 13$.  Solutions 
in each coordinate were iterated until no $3\,\sigma$ residuals remained.  
After convergence, the numbers of stars used averaged 350 in each coordinate,
and the unit weight errors averaged 6.9 and 7.3\masyr\ in 
$\Delta\mu_\alpha$ and $\Delta\mu_\delta$, respectively.  
The linear magnitude terms ($a_1, b_1$) averaged 
$(+1.3, +2.3)$~\masyr~mag$^{-1}$,
and typically were 3--4\,$\sigma$ significant in each field.
The quadratic magnitude terms ($a_2, b_2$) averaged 
$(-0.4, -0.5)$~\masyr~mag$^{-2}$,
and typically were 2\,$\sigma$ significant in each field.

Adding the $\Delta\mu_\alpha$, $\Delta\mu_\delta$ corrections (Eqs.~11a,b)
to each bright star's relative proper motion is equivalent to correcting the
linear ($a_1, b_1$) and quadratic ($a_2, b_2$) magnitude equations to a 
fiducial magnitude, $B = 12$, and then applying zero-point corrections
($a_0, b_0$) from relative to absolute proper motion.  

The fiducial magnitude $B = 12$ is where the NPM photography optimized the 
I\,(gr) and II\,(0) images.  In our view, the magnitude equations reflect
the circumstance that on the NPM blue plates, the overexposed I\,(0) image 
progressively encroaches on the grating and short-exposure images as
$B \rightarrow 8$, the limit of useful measurement \citep{dum78}.
For this reason, the magnitude-dependent part of the corrections determined 
here applies only to bright stars, not to the faint ($B \gtrsim 14$) stars 
whose isolated I\,(0) images were used.  

The zero-point corrections ($a_0, b_0$) defined at the fiducial magnitude 
$B = 12$, and derived from bright stars, apply to the faint stars as well,
because the bridge reductions (Section~\ref{BRG}) put all stars on the I\,(0) 
system.  In this view, $a_0, b_0$ simply reflect the mean motions of the faint 
($B \sim 16$) reference stars on the Tycho-2 system.

So, distinguishing ``bright'' and ``faint'' stars according to what image 
systems were measured, the corrections for bright stars are 
\begin{equation}
\mu_\alpha(abs)\ =\ \mu_\alpha(rel) + \Delta\mu_\alpha\ \ \ \ {\rm and} \ \ \ \
\mu_\delta(abs)\ =\ \mu_\delta(rel) + \Delta\mu_\delta
\end{equation}
\noindent
and for faint stars, we have
\begin{equation}
\mu_\alpha(abs)\ =\ \mu_\alpha(rel) + a_0\ \ \ \ {\rm and} \ \ \ \
\mu_\delta(abs)\ =\ \mu_\delta(rel) + b_0.
\end{equation}
These corrections put all the stars onto an absolute system, as defined
by the Tycho-2 stars.  The precision of the $\Delta\mu$ corrections in an
NPM2 field is about 0.5\masyr, the RMS size of the errors of $a_0$ and
$b_0$ from Eqs.~11.

\subsection{Accuracy of the Absolute Proper Motions} \label{PMACC}

The internal (measurement) errors of the NPM2 proper motions are 
$\sim$~5.6\masyr\ (Section~\ref{PMSC}).  Comparison of multiple measures
from field overlaps (Section~\ref{ERR}) estimates the RMS external accuracy 
of an individual proper motion to be 5.9\masyr.

Many of the intended uses of the NPM2 proper motions in stellar kinematics 
and galactic astronomy essentially involve averaging the motions of large
numbers of stars. In this context, the following question arises:

Without directly linking NPM2 to Hipparcos, how closely can we say that NPM2
is on the Hipparcos (ICRS) proper motion system?  Tycho-2 is nominally on the 
Hipparcos system.
From direct comparisons, using all 119,740 stars in common to the two catalogs, 
\citet{hog00a} concluded that the systematic errors of the Tycho-2 proper 
motions are $\lesssim 0.5$\masyr\ on angular scales of $6\degr$ or more 
(the size of the Lick plates, coincidentally), making the conservative 
assumption that Hipparcos' systematic errors are negligible.
\citet{hog00a} could not rule out that larger differences might exist on 
smaller angular scales, such as the $2\degr \times 2\degr$ size of the 
Astrographic Catalogue plates.  Presumably this is because of small-number 
statistics; the number of stars to compare averages just 3~deg$^{-2}$.
In any event, for NPM2 the relevant angular scale is that of the Lick plates.  

As a test for NPM2, we compared the Tycho-2 and Hipparcos proper motions, 
field by field, in the 347 NPM2 fields.  This was done twice: first, for all 
stars in common and second, in the limited magnitude range $9 < B <13$ used by 
NPM2.  Outliers with $\vert\Delta\mu\vert > 10$\masyr\ in either coordinate
were trimmed in order to restrict the comparisons to the most accurate data.

In the all-star comparison (30,517 stars; average 88 per field), the RMS values
of $\langle\Delta\mu_\alpha\rangle$ and $\langle\Delta\mu_\delta\rangle$
were 0.18 and 0.17\masyr, only $\sim 3$\% larger than the internal 
($\sigma/\sqrt N$) errors.  Only one of 347 NPM2 fields had 
$\langle\Delta\mu\rangle$ larger than 0.5\masyr\ in either coordinate.  
The \citet{hog00a} limit is quite conservative, being in effect a $3\,\sigma$ 
limit.  

In the faint-star comparison (14,323 stars; average 41 per field), 
the RMS values of $\langle\Delta\mu\rangle$ were substantially larger, 
0.28 and 0.30\masyr.  This reflects the lower precision of the $\Delta\mu$ 
values for stars this faint \citep[][Table~1]{hog00a}.  Again, the
$\langle\Delta\mu\rangle$ values were about the size of the internal errors.
None of 347 NPM2 fields had $\langle\Delta\mu\rangle$ larger than 1.0\masyr\ 
in either coordinate.

The main result of these proper motion comparisons is that, on the
$6\degr \times 6\degr$ angular scale of the NPM fields, any differences 
between the Tycho-2 and Hipparcos proper motion systems are in effect
too small to measure.  Formally, the external/internal error ratio (1.03)
in the all-star comparison suggests that systematics on the order of 
0.05\masyr\ (RMS) may be present, but this result is far from exact.

For NPM2, the RMS dispersion 0.3\masyr\ from the faint-star comparisons
is the relevant limit on how closely our reduction of the NPM2 proper motions 
to Tycho-2 is the equivalent of reducing them to the Hipparcos system.
In practice, for a given NPM2 field, this adds quadratically to the 0.5\masyr\
error of the NPM-Tycho corrections.  So, the NPM2 proper motions are on the 
Hipparcos (ICRS) system to an accuracy of 0.6\masyr.

\section{Photometry} \label{PHOT}

In each NPM2 field, the PMM instrumental magnitudes $m_b$ and $m_y$ are 
reduced into $B$~magnitudes and $B-V$ colors, with multiple measures for each 
star combined into a single magnitude and color.  Our experience with NPM1
and our tests for NPM2 (see Section~\ref{ERR}) show that the RMS errors of the
NPM photographic photometry are about 0.2~mag in $B$ and the same in $B-V$.
Color errors for very faint, red, or blue stars can be substantially larger.
We caution readers of this paper, and users of the NPM Catalogs, that the Lick 
astrograph plates are not suitable for more accurate photometry.  The NPM1 and 
NPM2 magnitudes and colors are accurate enough for statistical discussion of 
the proper motions, which is their intended purpose. 

Rather than describing the NPM2 photometric reductions in detail, 
we will give a broad outline, emphasizing several systematic improvements 
over the NPM1 procedures described in Paper~I:

\begin{enumerate}

\item
On each plate (B1, B2, Y2), we reduce the bright stars' ($B \lesssim  14$) 
long-exposure grating image photometry into the short-exposure II\,(0) system.
In each field, we reduce every star's B1 photometry into the system of the 
B2 plate.  For further reductions, $m_b$ is taken as the mean of the B1 and B2 
measures (both in the B2 system).  No photometric standards are needed for 
these differential reductions.\footnote{For historical accuracy, we note that
these ``photometric bridge reductions'' were actually used in the 1993 NPM1 
Catalog reductions, but had not been devised at the time Paper~I \citep{kjh87}
was written.}  Plate-constant models use six terms: zero point, $m_b$~or~$m_y$,
$X$, $Y$, $X^2$, and $Y^2$.  

\item
We use the Tycho-2 photometry \citep[$B_T,V_T$, reduced to Johnson $B,V$ 
using Eqs.~1.3.20 of][]{esa97}, as supplementary standards.  This gives an
order-of-magnitude increase over the number of primary standard stars 
(ICSS stars with BV photometry; Table~\ref{tbl-1}) available per field.

\item
For faint stars ($B \gtrsim 14$), measured as I\,(0) images, we derive $B$ and
$V$ magnitudes after differentially calibrating the $\sim$4~mag long$-$short 
exposure difference on each plate, effectively extending the Tycho-2 
photometric standards 4~mag fainter.  A simple plate model (zero point and 
linear magnitude term) usually suffices for this.  Separate solutions are done
in each color.  Typically, 20 to 30 faint photometric standard stars are used.  

\item
Plate models transforming $m_b$ into $B$, and $m_y$ into $V$ use six terms:
zero point, $m$, $m^2$, and field terms $X$, $Y$, and $R^2 = X^2 + Y^2$.  
Internal errors average 0.25~mag for bright stars, 0.30~mag for faint stars.
These values overestimate the true errors (Section~\ref{ERR}) of the NPM2 
photometry, because of the limited accuracy of the photometric standards.

\item
For bright stars, the $B-V$ color is directly calculated 
from $B$ and $V$.  For faint stars, $B-V$ is also determined from the
instrumental color $m_b - m_y$, using the statistical method used for NPM1
(Section VIIIe of Paper~I); the color adopted for NPM2 is generally the
mean of the two methods.  For $\sim$1.5\% of NPM2 stars, $B-V$ is not
determined, usually because the yellow plate measure is lacking.

\item
For fainter Tycho-2 stars ($11 \lesssim B \lesssim 14$), the precision of the 
Tycho-2 photometry \citep[see Table 2 of][]{hog00b} is comparable to NPM2.
For brighter stars, the Tycho-2 errors are much smaller than NPM2.
Consequently, for all Tycho-2 stars in the NPM2 catalog, we adopt the 
error-weighted mean of the Tycho-2 and NPM2 $B$ magnitudes and $B-V$ colors. 

\end{enumerate}

\section{Lick NPM2 Catalog} \label{CAT}

In this Section we outline the compilation and final content of the Lick NPM2
Catalog, determine the external errors of the NPM2 positions, proper motions,
and photometry, and compare these to the results achieved for NPM1.  
We also describe two important supplements to the NPM2 Catalog: the NPM2 
Cross-Identifications list and Klemola's ICSS Star Identification Catalog.

\subsection{Catalog Compilation} \label{COMP}

The NPM2 data reductions yielded 347 partially overlapping ``field catalogs'',
with a raw total of 288,615 stars (average $\sim$830 per field).  Following 
the procedures used in the production of the NPM1 Catalog \citep{han97},
the NPM2 Catalog was compiled by merging and combining data from the field 
catalogs.  First, each position was updated from its mean plate epoch 
($\sim$1968) to the common catalog epoch 2000, by applying the NPM2 proper 
motions. Then the stars were sorted into a one-degree zone-catalog format, 
with multiple measures (up to four per star) from the field overlaps 
($\sim 1^\circ$) averaged to give one entry per star, as in NPM1.  
No block adjustments were done, because of the need to preserve the 
statistical independence of each field, which is important for galactic 
and astrometric applications of the NPM data.  
The field overlaps were used to estimate the external errors 
of the NPM2 positions, proper motions, magnitudes, and colors, as well as to 
reject the small fraction ($ < 0.1$\%) of erroneous measurements not 
previously detected in the plate reductions, and to ``flag'' stars with 
discordant astrometry or photometry.

The completed Lick NPM2 Catalog \citep{hkjm03}, contains 232,062 stars,
a reduction of 20\% from the raw total of the 347 field catalogs.  
The catalog, in 108 one-degree declination zone files from $+83\degr$ 
to  $-23\degr$, was released on our WWW site\footnote{
{\tt http://www.ucolick.org/$\thicksim$npm/NPM2}}
in May 2003, along with extensive documentation, including error estimates
and file format details, in an accompanying ``ReadMe'' file.
The NPM2 Catalog is also available in the form of one 26~Mb file 
(compressed to 6.4~Mb) concatenating the 108 zones.  This version was also
deposited at the CDS Strasbourg data center (catalog number I/283A), 
where it is available by ftp and accessible by VizieR query.  

Table~\ref{tbl-3} illustrates the format and content of the NPM2 Catalog.
Following the convention of the NPM1 Catalog, each star is indexed by an NPM2
``name'' (e.g. +83.0001) reflecting the declination zone and a running number 
in J2000 right ascension order within the zone.  Each star's entry includes 
the absolute proper motion ($\mu_\alpha,\mu_\delta$) and $B$ magnitude.  
For 98.5\% of the stars the $B-V$ color is also given.  Other data given 
for each star are: the original mean epoch, a stellar class code, 
the number of NPM fields ($N_F$) on which the star was measured, 
and discrepancy flags for position, proper motion, and photometry.  

The NPM2 Catalog also lists identifications for 90,690 Tycho-2 stars, 20,426 
stars from the ACRS catalog, and 8,437 Hipparcos stars.  These categories 
overlap greatly.  The positional catalog stars total 91,648; nearly all have 
$B < 14$.  Non-catalog NPM2 stars are mostly faint ($B > 14$); 122,806 are 
anonymous stars chosen for astrometric and galactic studies.  The NPM2 Catalog
contains 34,868 ``special stars'' from Klemola's ICSS; about half of these 
have $B < 13$, and overlap substantially with the catalog stars. 
Identifications for the NPM2 ``special stars''
are discussed in Section~\ref{CROSS}.

\subsection{External Errors} \label{ERR}

Because the NPM fields (Figure~\ref{skyNPM}) overlap by at least $1\degr$ in 
each direction, the NPM2 catalog has 45,143 stars with multiple measures in 
overlapping NPM2 fields (37,854 stars have $N_F = 2$; 4,201 have $N_F = 3$;
and 3,088 have $N_F = 4$).
Another 2,642 NPM2 stars overlap with NPM1 at the edges of the Milky Way.
Because the NPM fields were measured and reduced separately, the multiple
measures from overlapping fields are statistically independent, so that the
RMS dispersion of the 2, 3 or 4 multiple measures about the mean value used 
in the NPM2 Catalog can be used to estimate the external errors of the NPM2 
positions, proper motions, magnitudes, and colors.  

Table~\ref{tbl-4} compares the NPM2 external errors with the internal error
(precision) estimates derived in Sections~\ref{POS}, \ref{PMS}, and \ref{PHOT}.
We also compare the NPM2 errors with the NPM1 Catalog errors given by 
\citet{han97}.

For the NPM2 positions, the internal errors at the mean plate epoch 
($\sim$1968) are calculated from the RMS residuals for the three sets 
(B1, B2, Y2) of plate reductions (Section~\ref{PPCR}).  
The external errors of the NPM2 positions were derived twice.  At the NPM2 
Catalog epoch 2000, the positional errors are the quadratic sum (see Eq.~1) 
of the pure positional errors at the mean plate epoch and the 
proper motion error accumulated from the mean epoch to 2000.  To evaluate the 
pure positional errors, we did a second overlap comparison using NPM2 positions
computed for the epoch 1968.  (The same procedure was used for NPM1.) 
The external errors at epoch 1968 are only $\sim$10\% larger than the internal 
error estimates, and nearly a factor of two better than NPM1.

For the NPM2 proper motions, internal errors cannot be obtained from the 
plate solutions, because the residuals largely represent stellar motions, 
not measurement errors.  Table~\ref{tbl-4} lists the precision of an 
individual proper motion measurement as estimated in Section~\ref{PMSC},
assuming 2\micron\ PMM measurement errors at each epoch and a 28~yr mean epoch
difference $\langle$B2$-$B1$\rangle$.  The external errors from the field 
overlap comparisons are $\sim$10\% larger.  The NPM2 proper motion errors are
$\sim$10--20\% larger than NPM1; this may reflect a slightly higher precision
for the Lick AME measurements used in NPM1.

The precision of the NPM2 absolute zero point errors is estimated by 
the RMS dispersion of the mean proper motion differences 
$\langle$Tycho-2$-$NPM2$\rangle$ for 347 NPM2 fields (Section~\ref{PMABS}).
The external error listed in Table~\ref{tbl-4} is the value derived in
Section~\ref{PMACC} for the accuracy to which each NPM2 field has been put 
onto the ICRS proper motion system.  These values are 3--4 times better than 
the zero point error in a typical NPM1 field, where $\sim$80 galaxies were
used to correct the proper motions to absolute.

For the NPM2 photometry, the RMS residuals from NPM2 plate reductions 
somewhat overestimate the internal errors because of the limited accuracy 
of the photometric standards.  In Table~\ref{tbl-4} the internal error 
given for the NPM2 $B$~magnitudes is the lower end of the range found in 
Section~\ref{PHOT}.  From our experience in NPM1 and NPM2, we adopt the 
same value for the $B-V$ colors.  To obtain valid external errors for $B$ and 
$B-V$, these were evaluated from field overlap comparisons using a version of 
the NPM2 Catalog that did {\sl not} average in any Tycho-2 photometry.
The photometric errors derived here apply to (generally faint) non-Tycho-2
stars in the NPM2 Catalog.  Overall, the NPM2 PMM photometry is roughly as 
accurate as NPM1's AME photometry.

Overlap comparisons were also done using 2,642 stars in common between the 
NPM1 and NPM2 Catalogs.  Mean and RMS differences were calculated after 
trimming 5\% of the outliers from each tail of the distribution.  Positions 
were compared at the epoch 1968.  The RMS differences were 210~mas in 
$\Delta\alpha$ and 230~mas in $\Delta\delta$.  These are 20--30\% larger than 
expected from the errors listed in Table~\ref{tbl-4}.  Positional zero point 
errors of order 100~mas in the individual NPM1 fields are the likely cause.
The RMS proper motion differences were 7.0\masyr\ in $\Delta\mu_\alpha$ and
8.0\masyr\ in $\Delta\mu_\delta$.  RMS photometry differences were 0.27~mag 
in $B$ and 0.26~mag in $B-V$ (2,478 stars).  The proper motion and photometry 
differences are consistent with the NPM1 and NPM2 errors in Table~\ref{tbl-4}.

We note that the NPM2 and NPM1--NPM2 field overlap error estimates were all 
derived using stars within $\sim$1$\degr$ of the edges of the $6\degr \times 
6\degr$ Lick plates, and may overestimate the errors of the majority of stars,
which are closer to the plate centers, where the image quality is better.
Also, the errors listed in Table~\ref{tbl-4} pertain to a single measurement
on one NPM2 field.  The data listed in the NPM2 Catalog for the 45,143 stars 
with $N_F > 1$ are means of $N_F =$ 2, 3, or 4 measures, and their actual
errors should be $N_F^{-0.5}$ times smaller, in the absence of systematic 
errors.  To be conservative, we chose to cite the single-measurement errors 
in Table~\ref{tbl-4} to represent the overall accuracy of the NPM2 data.
Finally, we note that the astrometric errors for stars brighter ($B < 10$) 
or fainter ($B > 17$) than the optimal range of measurement on the Lick plates
tend to be somewhat larger than the values listed in Table~\ref{tbl-4}.

\subsection{NPM2 Cross-Identifications} \label{CROSS}

To assist users of the NPM2 Catalog, we have compiled a supplementary 
catalog, the NPM2 Cross-Identifications and Appendices 
\citep[NPM2 Cross-IDs;][]{khjm04}, much as we did for NPM1 \citep{khj94}.
This catalog was released on our WWW site in March 2004, and will be
available at the CDS data center (catalog number I/293).  
The NPM2 Cross-IDs file, with 46,887 
entries, lists star names, stellar type classifications, and publication 
references for the 34,868 NPM2 ``special stars.'' (ICSS stars have multiple 
identities when chosen under more than one category of interest.)  Many of 
these identifications from the literature had been uncertain or ambiguous, 
and were verified by Klemola at the Lick Survey Machine, usually from finding
charts, variability, or proper motion.  After the NPM2 positions were 
available, Klemola re-confirmed many of these identifications using the 
CDS SIMBAD and NASA SkyView databases, as well as recent catalogs listing 
improved positions for variable stars, carbon stars, and other categories.
Based on this work, the NPM2 Cross-IDs file gives a reliability code for
each stellar identification.

Four Appendices to the NPM2 Catalog detail the 570 literature references and 
list the 164 stellar (occasionally, non-stellar) classes from which the NPM2
``special stars'' were selected.  A fifth Appendix provides 1,847 footnotes to
the NPM2 Catalog.  A ``ReadMe'' file briefly summarizes the NPM2 Cross-IDs and
provides file formats.  Another text file gives detailed documentation on the 
compilation and content of the NPM2 Cross-IDs, and discusses the Lick ICSS and
its relation to the NPM Catalogs.

The NPM2 Cross-Identifications will facilitate many practical uses of the 
NPM2 Catalog, allowing users to select ``special stars'' by category 
(e.g. RR~Lyrae variables, white dwarfs, faint blue stars, etc.) or from 
specific catalogs or literature sources, as well as to identify individual 
stars of interest.  We are using the NPM1 and NPM2 Cross-Identifications
files to produce combined NPM1~$+$~NPM2 ``mini-catalogs'' for use in galactic 
astronomy.  We intend to post these on our WWW site.
Major categories of interest include 
6000 OB stars, 
5100 faint blue stars/objects, 
4900 red giants, 
3900 emission-line stars, 
3500 high-proper-motion stars, 
2100 RR~Lyrae type variables, 
1600 white dwarf stars/candidates, 
1100 horizontal branch stars, and 
 900 carbon stars.  
Variable stars of all types total 12,500.
Halo population stars total about 5000.
(These counts are preliminary, and may change when individual categories 
are studied in detail.)

\subsection{ICSS Star Identification Catalog} \label{SIC}

For numerous reasons, Klemola's ICSS contains many stars that are not included 
in the NPM Catalogs.  The ICSS has been continually updated for completeness, 
but it was not practical for NPM1 to include stars from sources published 
after the plates were measured.  Few NPM2 fields were re-surveyed for 
additional stars.  Later additions to the ICSS could now be recovered from 
the PMM scans, but that is beyond the scope of the Lick program.  For both 
NPM1 and NPM2, many ICSS stars could not be reliably identified at the Survey 
Machine, chiefly because of poor positions or lack of finding charts.
In all, about 46\% of the 123,000 ICSS stars in the 347 NPM2 fields
were not selected in the plate surveys.  
For NPM2, another 14\% of the ICSS stars were identified at the Survey Machine
but were coded for rejection before the PMM star selection stage.  Very bright 
stars ($B \lesssim 9$), or stars too faint ($B \gtrsim 18$) for useful 
measurement were not generally accepted for NPM2.  Variable stars with large 
amplitude often did not have useful images at both epochs.  Many stars had
no unblended images to use, due to crowding in the dense low-latitude fields.
The loss fraction for ``special stars'' is much larger than for catalog or 
anonymous stars, where we had more freedom to reject problematic stars, 
and choose substitutes, on the PMM scans.  Another 4\% of the ICSS stars
were not selected on both the B1 and B2 PMM scans. These stars cannot be used
for NPM2.  Finally, 7\% of the ICSS stars were rejected at various stages in 
the NPM2 data reductions, when faint, blended or otherwise unsuitable images 
caused large residuals.  In all, the 34,868 (of 123,000) ICSS stars in the 
NPM2 Catalog represent a survival rate less than 30\%.  The high attrition of
``special stars'' in NPM2 is an unfortunate consequence of our conservative 
policy of minimizing spurious results by aggressively rejecting problematic 
stars.

Although many ICSS stars were ``lost'' to NPM2, Klemola's plate surveys contain
much information of value to observers and database compilers. Over 30,000 ICSS
stars not in NPM2 were positively identified at the Survey Machine, and their 
$X,Y$ coordinates were visually measured with $\sim$20~\micron\ precision, 
$\sim$1~arcsec on the NPM plates.  This often represents an order-of-magnitude
improvement over the previous crude positions.  Klemola will compile a 
``Star Identification Catalog'' (SIC) with confirmed identifications and 
arcsecond-level positions for this subset of the ICSS.  The SIC data will be 
a valuable contribution to stellar databases such as SIMBAD, and will prove 
useful for identifying stars in existing and future databases, and facilitating
further observations for many stars of particular astronomical or 
astrophysical interest.  

\section{Summary and Discussion} \label{SUM}

The two NPM Catalogs total 378,360 stars (148,940 in NPM1 plus 232,062 in NPM2,
less 2,642 in both due to plate overlaps) from $8 \lesssim B \lesssim 18$,
covering the northern two-thirds of the sky ($\delta \gtrsim -23\degr$). 
NPM2 covers only $\sim$40\% of the sky area of NPM1, but its star density is 
four times as high. 
This partly reflects that many classes of stars are strongly concentrated
towards the galactic plane.  A particular scientific value of NPM2 lies in its
low-latitude (\,$|b| < \ \sim$10$\degr$) sky coverage.  Important tracers of
the galactic disk and spiral arms (Cepheids, OB stars) are generally found only
at low latitudes.  Other classes of stars in the Lick Input Catalog of Special
Stars that are chiefly represented at low latitudes are carbon stars, red
giants, and many types of variable stars.  Even halo population stars such as
the RR~Lyrae variables are found in large numbers at lower galactic latitudes
in the NPM survey, because of their concentration toward the galactic center.
Many of the 34,868 NPM2 ``special stars'' are too distant, and thus too faint,
to be represented in significant numbers in the Hipparcos or Tycho Catalogues.
The NPM2 Catalog also includes two large, kinematically unbiased, sets of stars
useful for studies of galactic structure and kinematics: 91,648 positional 
catalog stars with $B \lesssim 14$; and 122,806 faint ($B \gtrsim  14$)
anonymous stars.

Because the long-term value of the NPM proper motions depends critically on 
their precision and accuracy, great effort was devoted to testing their
quality, with the goal of calibrating and eliminating systematic errors.  
Repeated tests confirmed the high precision and accuracy of the NPM1 proper 
motions \citep{han87, han97, plx98b}.  
As detailed in Table~\ref{tbl-4}, this level of quality has been met or 
exceeded for NPM2.  The RMS errors over the magnitude range $10 < B < 17$ 
are 5--6\masyr\ for both catalogs.  The absolute zero point of 
proper motion, determined independently in each NPM field, is accurate to 
2\masyr\ for NPM1, and 0.6\masyr\ for NPM2.

The high quality of the NPM proper motions has also been demonstrated in 
their applications to several problems in galactic structure and astrometry, 
such as the solar motion and galactic rotation \citep{han87}, 
the absolute motions of 14 globular clusters \citep{ch93}, 
the statistical parallax luminosity calibration of 200 RR~Lyrae stars 
\citep{lay96}, and the linking of Hipparcos to the inertial proper 
motion system \citep{plx98b}.  In particular, the latter two results confirm
that the absolute proper motion zero points for the NPM 
``bright'' ($B \sim 12$) and 
``faint''  ($B \sim 16$) stars are consistent to better than 1\masyr.

The NPM2 positions at 1968, near the mean epoch of observation, are twice as
precise as NPM1 (RMS error 80~mas for NPM2, 150~mas for NPM1), and their 
systematic accuracy is ten times better (10~mas for NPM2, 100~mas for NPM1).
These improvements are due to the use of a much better plate model, and an 
order of magnitude more reference stars per field, than NPM1.  The NPM2 
photographic positions, though not having the high precision of modern
positional catalogs, are in fact of comparable or better quality than Tycho-2
for $V \geq 11.5$ \citep[see Table~2 and Figure~2 of][]{hog00b}.
Likewise, the RMS errors ($\sim$0.2~mag) of the NPM2 photographic photometry,
which was chiefly intended for statistical use, equal or surpass Tycho-2 for 
$V \geq 12$.

The publication of the NPM2 Catalog and Cross-IDs in 2003-4 completes the Lick
Northern Proper Motion program as it was originally conceived, but it does not
mean that our work is concluded.  Much can be done to apply the NPM data to 
important research problems in galactic structure, stellar kinematics, and
astrometry.  Applications now underway at Lick include: (1)~extension of the 
NPM1 \citep{han87} galactic rotation and solar motion study to cover the sky 
north of $-23\degr$ and to determine $dV /d\vert Z\vert$, the progressive lag
behind circular rotation with distance from the galactic plane, due to the 
changing mix of stellar populations with different scale heights; and 
(2)~use of the NPM1 and NPM2 Cross-IDs, and the method of reduced proper 
motions, for a comprehensive identification of $\sim$5,000 halo population 
stars in the two catalogs.  

Many other uses of the NPM1 and NPM2 data can be foreseen.  The two Lick NPM 
Catalogs uniquely represent a small, but carefully chosen and rigorously 
reduced, subset of stars in the northern two-thirds of the sky, intended
to serve as a database for galactic research needing accurate proper motions
for stars selected with little or no kinematic bias, and to supply proper 
motions for well-identified stars in categories of high astronomical interest.
The PMM scans of all 1,246 NPM fields, containing data for several hundred 
million stars, are another important product of the NPM program, which can
be used to expand the NPM Catalogs, much as the Yale SPM program is doing in 
the southern sky, or to incorporate data from the Lick plate epochs into 
modern all-sky proper motion catalogs.

\acknowledgements

The Northern Proper Motion program represents a half-century of work by three 
generations of astronomers at Lick Observatory, whose contributions were
acknowledged in some detail in Paper~I.  Once again, we thank W.H.~Wright,
C.D.~Shane, C.A.~Wirtanen, S.~Vasilevskis, and E.A.~Harlan for the foresight,
innovation, and dedication that brought the NPM program to the point where
we could successfully complete it.
We also thank the National Science Foundation for its long-term support of 
the Lick NPM program.  The NPM2 phase was supported by NSF grants AST-9530632 
and AST-9988105.  We thank the Yale Southern Proper Motion group 
(W.F.~van~Altena, I.~Platais, and T.M.~Girard) for their help in developing 
software to process the PMM plate scans, and we thank the staff members
of the US Naval Observatory Flagstaff Station who scanned nearly 4,000 NPM
plates on the PMM.  We thank Norbert Zacharias (USNO) for making the 
NPM2~$-$~UCAC2 positional comparison.  We thank Imants Platais for his 
detailed, thoughtful, and prompt referee's report.  This research has made use
of the SIMBAD database, operated at CDS, Strasbourg, France.

\clearpage
\begin{deluxetable}{lccc}
\tablenum{1}
\tablecolumns{4}
\tablewidth{0pc}
\tablecaption{Content of the NPM1 and NPM2 Catalogs 
(stars per field)\label{tbl-1}}
\tablehead{ 
\colhead{Star class} & 
\colhead{Mag. range} & 
\colhead{NPM1} & 
\colhead{NPM2}\\
\colhead{ } &\colhead{(approx.)} &\colhead{899 fields} &\colhead{347 fields}
}
\startdata 
Faint anonymous stars (FAS) &  ($15 < B < 17$) & 70--140       & $\sim$400         \\
Bright anonymous stars      &  ($10 < B < 13$) & 15--30        & \nodata           \\
Positional reference stars  &  ($ 9 < B < 12$) & $\sim$50 ACRS\tablenotemark{a} 
& $\sim$400 Tycho-2 \\
Photometric standard stars  &  ($ 8 < B < 18$) & 0--100        & 0--100            \\
ICSS ``special stars''      &  ($ 8 < B < 18$) & $\sim$30      & $\sim$100         \\
Other catalog stars         &  ($ 8 < B < 14$) & \nodata       & $\sim$40 Hipparcos\\
\enddata
\tablenotetext{a}{NPM1 used ACRS catalog data for reference stars 
originally selected from the AGK3 ($\delta > -2\fdg5$) or SAO 
($\delta < -2\fdg5$) catalogs.}
\end{deluxetable}

\clearpage
\begin{deluxetable}{cccrcccccccccccccc}
\tablenum{2}
\tablecolumns{18}
\tabletypesize{\scriptsize}
\tablewidth{0pc}
\tablecaption{
Lick Astrograph Plate Constant Model\tablenotemark{a}\label{tbl-2}
}
\tablehead{ 
\colhead{} &
\colhead{} &
\colhead{} &
\colhead{} &
\colhead{} &
\colhead{} &
\colhead{} &
\colhead{} &
\multicolumn{4}{c}{Quadratic\tablenotemark{d}} &
\colhead{} &
\multicolumn{5}{c}{Cubic\tablenotemark{e,{\rm f} }}\\
\colhead{Term} & 
\colhead{Scale\tablenotemark{b}} & 
\multicolumn{2}{c}{Rot\tablenotemark{c}} & &
\multicolumn{2}{c}{Zero Pt} & &
\colhead{$P_X$} & 
\colhead{$Q_X$} & 
\colhead{$P_Y$} & 
\colhead{$Q_Y$} & &
\colhead{$D_{1X}$} & 
\colhead{$D_{2X}$} & 
\colhead{$D_{1Y}$} & 
\colhead{$D_{2Y}$} & 
\colhead{$D_{3}$}
}
\startdata
$c_i$ & $c_1$ & $c_2$ & $c_3$ & & $c_4$ & $c_5$ & & $c_6$ & $c_7$ & $c_8$ 
& $c_9$ & & $c_{10}$ & $c_{11}$ & $c_{12}$ & $c_{13}$ & $c_{14}$\\
$T_{i,X}$ & $X$   & $Y$   & 0     & & 1     & 0     & & $X^2$ & $XY$  & 0     
& 0     & & $X^3$    & $XY^2$   & 0        &   0      & $X^2Y$  \\
$T_{i,Y}$ & $Y$   & $0$   & $-X$  & & 0     & 1     & & 0     & 0     & $XY$  
& $Y^2$ & & 0        & 0        & $X^2Y$   & $Y^3$    & $XY^2$  \\
\enddata
\tablenotetext{a}{For NPM2, the same model is used for positions {\sl and} 
proper motions.  The plate center is the origin of the $X,Y$ coordinates.}
\tablenotetext{b}{One scale term ties the $X$ and $Y$ solutions together.}
\tablenotetext{c}{The usual rotation term is split into $X,Y$ components.}
\tablenotetext{d}{The usual plate tilt terms $P,Q$ are split into $X,Y$ 
components.}
\tablenotetext{e}{The usual cubic distortion term $\ X(X^2+Y^2)\ $ in $X$,
$\ Y(X^2+Y^2)$ in $Y\ $ is split into four parts.}
\tablenotetext{f}{The anomalous distortion term $D_3$ is the same in 
$X$ and $Y$.}
\end{deluxetable}

\clearpage
\begin{deluxetable}{rrrrrrr}
\tablenum{3}
\tablecolumns{7}
\tabletypesize{\scriptsize}
\tablewidth{0pc}
\tablecaption{
Lick NPM2 Catalog: Declination Zone $+83\degr$\label{tbl-3}
}
\tablehead{ 
\colhead{~~NPM2~~~~~~~R.A~~~(J2000)~~~Dec.~~~~~} &
\colhead{$~\mu_\alpha$} &
\colhead{$~\mu_\delta$} &
\colhead{~~$B$~~~~$B$-$V$} &
\colhead{Epoch~~~Codes~~} &
\colhead{Tycho-2} &
\colhead{ACRS~~~~Hipp~~~}
}
\startdata
+83.0001~ 21 35 05.079~ +83 04 23.46 &   0.77 & --0.43 & 11.60~ 1.57 &  
69.10 ~~2 1 000 & 4649 02347 1 & 562334~~ 000000\\
+83.0002~ 21 44 18.741~ +83 04 42.14 &   1.61 & --0.96 & 15.11~ 0.84 &  
69.10 ~~1 1 000 & 0000 00000 0 & 000000~~ 000000\\
+83.0004~ 21 47 19.602~ +83 01 56.60 & --3.33 & --0.41 & 11.12~ 1.44 &  
69.10 ~~2 1 000 & 4649 02429 1 & 000000~~ 000000\\
+83.0005~ 21 51 14.098~ +83 11 32.50 &   0.13 & --0.05 & 10.55~ 0.47 &  
69.10 ~~2 1 000 & 4649 01903 1 & 563517~~ 000000\\
+83.0006~ 21 53 52.852~ +83 02 15.28 & --0.48 & --1.03 & 16.10~ 0.59 &  
69.10 ~~1 1 000 & 0000 00000 0 & 000000~~ 000000\\
+83.0007~ 22 04 29.183~ +83 04 25.78 & --0.16 &   0.04 & 15.37~ 0.81 &  
69.10 ~~1 1 000 & 0000 00000 0 & 000000~~ 000000\\
+83.0008~ 22 06 21.215~ +83 04 57.76 &   2.49 & --3.03 & 12.31~ 0.82 &  
69.10 ~~2 1 000 & 4650 02792 1 & 000000~~ 000000\\
+83.0009~ 22 14 06.989~ +83 11 11.09 & --0.22 & --0.88 & 16.63~ 1.01 &  
69.10 ~~1 1 000 & 0000 00000 0 & 000000~~ 000000\\
+83.0010~ 22 16 43.463~ +83 01 29.07 & --0.69 &   1.34 & 11.53~ 1.33 &  
69.10 ~~2 1 000 & 4650 02484 1 & 565099~~ 000000\\
+83.0011~ 22 24 04.573~ +83 11 10.13 & --0.20 &   0.86 & 15.67~ 0.88 &  
69.10 ~~1 1 000 & 0000 00000 0 & 000000~~ 000000\\
+83.0012~ 22 28 05.231~ +83 10 31.25 & --0.98 & --1.22 & 10.95~ 0.48 &  
69.10 ~~2 1 000 & 4650 00738 1 & 565757~~ 000000\\
+83.0013~ 22 33 51.856~ +83 07 01.28 & --1.23 &   0.28 & 14.95~ 0.95 &  
69.10 ~~1 1 000 & 0000 00000 0 & 000000~~ 000000\\
+83.0014~ 22 34 41.628~ +83 08 24.50 & --2.53 & --2.01 & 12.11~ 0.67 &  
69.10 ~~2 1 000 & 4650 01631 1 & 000000~~ 000000\\
+83.0015~ 22 38 16.266~ +83 00 50.42 & --0.64 &   1.37 & 11.90~ 1.10 &  
69.10 ~~2 1 000 & 4650 00230 1 & 566296~~ 000000\\
+83.0016~ 22 38 33.253~ +83 10 00.24 &  25.75 &  17.04 & 14.44~ 1.57 &  
69.10 ~~4 1 000 & 0000 00000 0 & 000000~~ 000000\\
+83.0017~ 22 44 21.540~ +83 10 08.08 & --0.26 &   0.02 & 15.63~ 1.53 &  
69.10 ~~1 1 000 & 0000 00000 0 & 000000~~ 000000\\
+83.0018~ 22 52 42.965~ +83 18 54.85 &   4.12 &   4.16 & 12.07~ 0.38 &  
69.10 ~~2 1 000 & 4650 01471 1 & 000000~~ 000000\\
+83.0019~ 22 55 19.344~ +83 08 54.31 &   2.69 &   2.05 & 15.59~ 0.59 &  
69.10 ~~1 1 000 & 0000 00000 0 & 000000~~ 000000\\
+83.0020~ 22 57 04.531~ +83 03 10.02 &   2.22 &   1.06 &  9.31~ 1.06 &  
69.10 ~~2 1 000 & 4650 00610 1 & 000000~~ 113328\\
+83.0021~ 23 05 31.480~ +83 06 31.71 &   2.45 & --0.36 & 14.98~ 0.99 &  
69.10 ~~1 1 000 & 0000 00000 0 & 000000~~ 000000\\
+83.0022~ 23 18 07.412~ +83 00 38.89 & --0.67 &   0.36 & 11.00~ 1.39 &  
69.10 ~~6 1 000 & 4650 01027 1 & 568200~~ 000000\\
+83.0023~ 23 22 05.225~ +83 01 52.70 &   0.53 & --1.66 & 11.56~ 0.64 &  
69.10 ~~2 1 000 & 4650 01177 1 & 000000~~ 000000\\
+83.0024~ 23 31 57.330~ +83 05 08.36 &   0.00 & --0.36 & 11.23~ 1.68 &  
69.10 ~~2 1 000 & 4650 01108 1 & 568850~~ 000000\\
\enddata

\tablecomments{
Table~\ref{tbl-3} shows the first (and smallest) of the 108\, one-degree 
declination zone files ($+83\degr$ to $-23\degr$) which make up the 
232,062-star Lick NPM2 Catalog.  This table illustrates the catalog's format 
and content.  Zones average 2,150 stars; the largest has over 4,200.
The machine-readable catalog and full documentation are available at 
{\tt http://www.ucolick.org/$\sim$npm/NPM2} and at the CDS Strasbourg data 
center (catalog number I/283A).
Positions are in the J2000 coordinate system {\sl at the catalog epoch} 2000.0.
Proper motion units are arcsec~cent$^{-1}$; multiply by 10 for mas~yr$^{-1}$.
Magnitudes and colors are from NPM2 photographic photometry and the Tycho-2 
Catalogue, as explained in Section~6.  
``Epoch'' is the original mean plate epoch for each star.
``Codes'' are star class, number of fields, and error flags, as explained in 
the documentation.  Tycho-2, ACRS, and Hipparcos numbers are given for NPM2 
stars from those catalogs.  Other identifications, stellar classifications,
and literature references are given in a supplementary catalog,
the NPM2 Cross-Identifications.}
\end{deluxetable}

\clearpage
\begin{deluxetable}{lccccccc}
\tablenum{4}
\tablecolumns{8}
\tablewidth{0pc}
\tablecaption{NPM Internal and External Errors\tablenotemark{a}\label{tbl-4}}
\tablehead{ 

\colhead{Errors} & 
\multicolumn{2}{c}{Position~[mas]} & 
\colhead{ } &
\colhead{Proper Motion~[mas~yr$^{-1}$]}& 
\colhead{ } &
\multicolumn{2}{c}{Photometry~[mag]} \\
\colhead{ } &
\colhead{1968} &
\colhead{2000} &
\colhead{ } &
\colhead{RMS~~~~~Zero Point~} &
\colhead{ } &
\colhead{$B$ } &
\colhead{$B-V$ }
}
\startdata 
NPM2~~internal~~~ & ~~77 & \nodata & & 5.6~~~~~~~~~0.5~~~~ & & 0.25 & 0.25 \\
NPM2~~external~~~ & ~~84 & 206     & & 5.9~~~~~~~~~0.6~~~~ & & 0.18 & 0.20 \\
NPM1~~Catalog~~~  &  150 & 220     & & 5~~~~~~~~~~~2~~~~~~ & & 0.20 & 0.15 \\
\enddata

\tablenotetext{a}{For positions and proper motions, errors in $\alpha$ and 
$\delta$ components are nearly equal; listed values are the mean of the two 
components.}

\tablecomments{Internal errors represent RMS residuals from NPM2 plate 
reductions.  External errors are determined using over 45,000 stars 
in NPM2 field overlaps.  Positions are compared at 1968, near the mean plate 
epoch, and at 2000, the NPM2 Catalog epoch.  Proper motion RMS errors
are for individual stars; zero point errors are for individual fields.
NPM1 Catalog errors are external errors from Hanson (1997).}

\end{deluxetable}

\clearpage

\clearpage

\clearpage

\clearpage

\clearpage

\clearpage

\end{document}